\documentclass[%
 reprint,
 amsmath,amssymb,
 aps,
]{revtex4-2}

\usepackage{graphicx}
\usepackage{dcolumn}
\usepackage{bm}

\usepackage{physics}
\usepackage{natbib}
\graphicspath{{./}{figures/}}
\graphicspath{{./}{figures_amend/}}

\usepackage{xcolor}

\begin{document}

\preprint{APS/123-QED}

\title{Dissipation and particle acceleration in astrophysical jets with velocity and magnetic shear:\\Interaction of Kelvin-Helmholtz and Drift-Kink Instabilities}

\author{Tsun Hin Navin Tsung$^{1,3}$}
\email{tsunhinnavin.tsung@colorado.edu}
\author{Gregory R. Werner$^{1}$}
\author{Dmitri A. Uzdensky$^{1,2}$}
\author{Mitchell C. Begelman$^{3,4}$}
\affiliation{$^{1}$Center for Integrated Plasma Studies, Physics Department, 390 UCB, University of Colorado, Boulder, CO 80309, USA}
\affiliation{$^{2}$Rudolf Peierls Centre for Theoretical Physics, Clarendon Laboratory, University of Oxford, Parks Road, Oxford OX1 3PU, UK}
\affiliation{$^{3}$JILA, University of Colorado and National Institute of Standards and Technology, 440 UCB, Boulder, CO 80309-0440, USA}
\affiliation{$^{4}$Department of Astrophysical and Planetary Sciences, University of Colorado, 391 UCB, Boulder, CO 80309, USA}

\date{\today}

\begin{abstract}
We present 2D particle-in-cell simulations of a magnetized, collisionless, relativistic pair plasma subjected to combined velocity and magnetic-field shear, a scenario typical for astrophysical black-hole jet-wind boundaries. We create conditions where only the Kelvin-Helmholtz (KH) and Drift-Kink (DK) instabilities can develop, while tearing modes are forbidden. We find that DKI can effectively disrupt the cat's-eye vortices generated by~KHI, creating a turbulent shear layer on the DK timescale. This interplay leads to a significant enhancement of dissipation over cases with only velocity shear or only magnetic shear. Moreover, we observe efficient nonthermal particle acceleration caused by the alignment of the instability-driven electric fields with Speiser-like motion of particles close to the shear interface. This study highlights the sensitivity of dissipation to multiple simultaneous instabilities, thus providing a strong motivation for further studies of their nonlinear interaction at the kinetic level.
\end{abstract}

\maketitle


\emph{Introduction.}--Relativistic astrophysical jets are inferred to have a spine-sheath structure, i.e. a faster moving spine encompassed by a slower sheath \citep{Attridge_etal-1999,Aloy_etal-2000,Pushkarev_etal-2005,Gabuzda_etal-2014,Bruni_etal-2021}, implying velocity shears at the jet boundaries. Such shear is prone to the Kelvin-Helmholtz (KH) instability, which has been studied extensively using linear theory \cite{Blandford_Pringle-1976,Ferrari_etal-1978,Ferrari_etal-1980,Ferrari_Trussoni-1983,Fiedler_Jones-1984,Bodo_etal-2004,Osmanov_etal-2008,Prajapati_Chhajlani-2010,Sobacchi_Lyubarsky-2018,Chow_etal-2023}, fluid simulations \cite{Keppend_Toth-1999,Ryu_etal-2000,Zhang_etal-2009,Hamlin_Newman-2013,Perucho_etal-2004}, and kinetic simulations \cite{Alves_etal-2012,Liang_etal-2013,Nishikawa_etal-2014,Sironi_etal-2021,Meli_etal-2023}. These jets have also been observed to be magnetized, threaded either by a toroidal or helical magnetic field \cite{Kharb_etal-2009,Gabuzda-2018,Gabuzda-2021}. Current sheets (i.e. magnetic shear) are likely to be present in jets too due to changes in the field orientation and strength across the jets' cross-sections. Current sheets are susceptible to tearing modes/magnetic reconnection \cite{Lyubarsky-2005,Werner_Uzdensky-2021}, which create plasmoid chains \cite[e.g.,][]{Uzdensky_etal-2010} and are efficient particle accelerators \cite{Zenitani_Hoshino-2001,Zenitani_Hoshino-2005,Zenitani_Hoshino-2007,Zenitani_Hoshino-2008, Sironi_Spitkovsky-2014,Guo_etal-2014,Werner_etal-2016,Werner_Uzdensky-2021, Sironi_etal-2016}. 
They are also prone to Drift-Kink (DK) instability \cite{Zhu_Winglee-1996,Daughton-1999,Pritchett_etal-1996,Zenitani_Hoshino-2007,Zenitani_Hoshino-2008,Barkov_Komissarov-2016, Werner_Uzdensky-2021}, which grows faster during the linear stage than the tearing modes in a weak guide field, but is otherwise not as efficient a particle accelerator as reconnection \cite{Zenitani_Hoshino-2007}. How the instabilities arising from velocity and magnetic shear interact with one another is an open question that has only now begun to be explored. A practical question is how such interaction affects energy dissipation and nonthermal particle acceleration (NTPA). Sironi et al. \cite{Sironi_etal-2021} considered a 2D jet-wind model --- a `jet' medium made up of pair plasma and a `wind' medium made up of normal (electron-ion) plasma with a relativistic velocity shear between them, and magnetic field that was helical in the jet, and toroidal and significantly weaker in the wind. The study found that KH vortices can wrap the field lines over each other, creating current sheets which trigger reconnection and considerable dissipation, but the specific setup of that study precluded the authors from exploring the effects of~DKI. 

\begin{figure}
    \centering
    \includegraphics[width=0.45\textwidth]{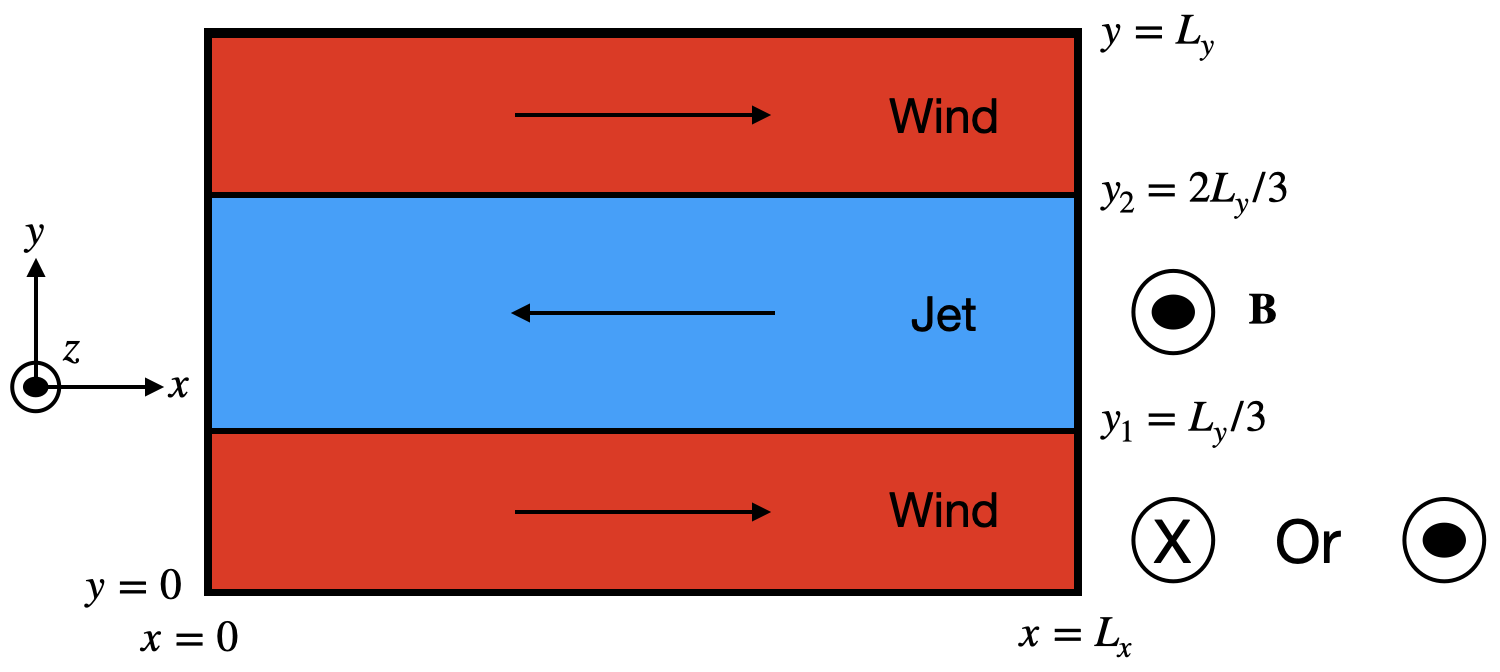}
    \caption{Setup for the jet-wind model.
    }
    \label{fig:PIC_setup}
\end{figure}

In this Letter we consider a similar computational setup (see fig.\ref{fig:PIC_setup}), but simplified in order to explore the joint effects of KH and DK modes, and of their nonlinear interplay, on  magnetic and bulk-kinetic dissipation. Here for the first time we incorporate the effects of velocity and magnetic shear, which create conditions in which DK coexisits with~KH, while tearing is forbidden. Using first-principles particle-in-cell (PIC) simulations, we show that the interaction between the KH and DK instabilities creates qualitatively new structures and very different amounts of dissipation compared to the case when only one instability is present. It leads also to nonthermal particle acceleration with power laws that are different. Thus, dissipation is highly sensitive to the interplay of the two instabilities.

\emph{Methods.}--We perform relativistic 2D electromagnetic PIC  simulations using the code {\sc Zeltron}~\cite{Cerutti_etal-2013}. We consider a relativistically warm ($\theta \equiv T/m_e c^2\sim1$) electron-positron pair plasma in a double layer shear flow (fig.\ref{fig:PIC_setup}). 
Our setup consists of two regions, the `jet' (blue) and the `wind' (red), with two thin interface layers between them.
In this setup, $x$ is the direction of the flow and $y$ is the direction of variation, representing, respectively, the axial and radial directions in a jet. We impose an initial out-of-plane magnetic field that is perpendicular to the flow and the gradient direction, i.e. $\vb{B} = B(y)\vu{z}$, mimicking a jet's toroidal field. We simulate the $x,y$-plane keeping all three components for vector quantities. Since the $z$-direction is not simulated, no tearing modes can be excited (a full 3D study is left to future work). This allows us to focus on the KH and DK modes and their interplay. 
We run the simulations in the frame where the jet and the wind are moving at the same speed, but in opposite ($\pm \vu{x}$) directions. 

The setup is characterized by the following dimensionless parameters: the 4-velocity of the jet's bulk flow $u_j = \Gamma_j\beta_j$, the jet relativistic temperature $\theta_j= T_j/m_e c^2$, the jet magnetization $\sigma_j = B_j^2/4\pi w_j$ (where $w_j$ is the jet relativistic enthalpy density), and the ratio of wind to jet magnetic field $B_w/B_j$. The jet magnetization $\sigma_j$ and relativistic temperature $\theta_j$ are set to 1 initially, and the jet and wind total rest-frame densities (electron plus position) are initially equal. We connect the jet and wind velocities and magnetic fields smoothly by a tanh function with transition half-width~$\Delta$. 
Assuming pressure balance, Maxwell's equations, the ideal gas law, and that the initial electric field is given by the motional E-field $\vb{E} = -\vb{v}\cross\vb{B}/c$, the $y$-profiles of electric field, the electrons' and positrons' velocity, density and temperature can be determined. The velocities of the particles are then sampled from the local drifting Maxwell-J\"uttner (MJ) distribution. Details for this setup can be found in Appendix \ref{app:setup}. With this setup, we ran two control cases with only one type of shear present $(u_j, B_w/B_j) = (0.5,+1)$, and $(0, -1)$ (hereafter referred to as the VS and MS case respectively), and progressively vary $u_j:\{0.01,0.03,0.05,0.07,0.1,0.2,0.3,0.4,0.5,0.6,0.7,0.8,0.9\}$ keeping $B_w/B_j = -1$ to investigate the effects of velocity shear on DKI.

The simulation domain has periodic boundaries and spans $L_x = 100 d_{e,j}$ in the $x$-direction and $L_y = 300 d_{e,j}$ in the $y$-direction and is resolved fiducially by $1024\times 3072$ grid cells, which resolves the jet electron inertial length~$d_{e,j}$ by $\approx$10 cells. 
The initial half-thickness of the two shear layers, located at $y_{1,2} = \{L_y/3, 2L_y/3\} = \{100, 200\}\, d_{e,j}$, is set to $\Delta = 5 d_{e,j}$.
For relativistically warm ($\theta\sim 1$), moderately magnetized ($\sigma\sim 1$) plasma, the jet electron Debye length $\lambda_{\mathrm{D}e}$, the average gyroradius~$\rho_{e,j}$, and the electron inertial length $d_{e,j}$ are roughly equal, so all the kinetic scales are well resolved (exact definitions for~$d_{e,j},\Lambda_{\mathrm{D},e},\rho_{e,j}$ are given in Appendix~\ref{app:microscales}). We created 64 particles per cell (PPC) per species (totaling 128~PPC); this resulted in $d^2_{e,j}/\Delta x\Delta y\times \mathrm{PPC} \sim 6400$ particles per area spanned by the jet electron skin depth, sufficient for our purposes. We ran our simulations for approximately $50 L_x/c$ to ensure the saturated stage was reached. For kinetic diagnostic purposes, we randomly selected and tracked the trajectories of $10^5$ electrons and positions throughout the simulation.

The main focus of this study is on the evolution of the total magnetic and bulk kinetic energies. The total magnetic energy is calculated as $E_B \equiv \int B^2/8\pi\,\mathrm{d}V$. To calculate the total bulk kinetic energy, we first perform Lorentz transformation into the local zero-particle-flux (Eckart) frame at each cell, 
calculate the local pressure tensor there, and then subtract it from the total stress tensor to obtain the bulk-flow stress tensor. 
Integrating its trace over the box volume gives the total bulk kinetic energy $E_\mathrm{KE}$ (details of this procedure are given in Appendix \ref{app:decompose}). As a consistency check, the total energy in all our simulations is conserved to within~0.2\%.

\begin{figure}
    \centering
    \includegraphics[width=0.45\textwidth]{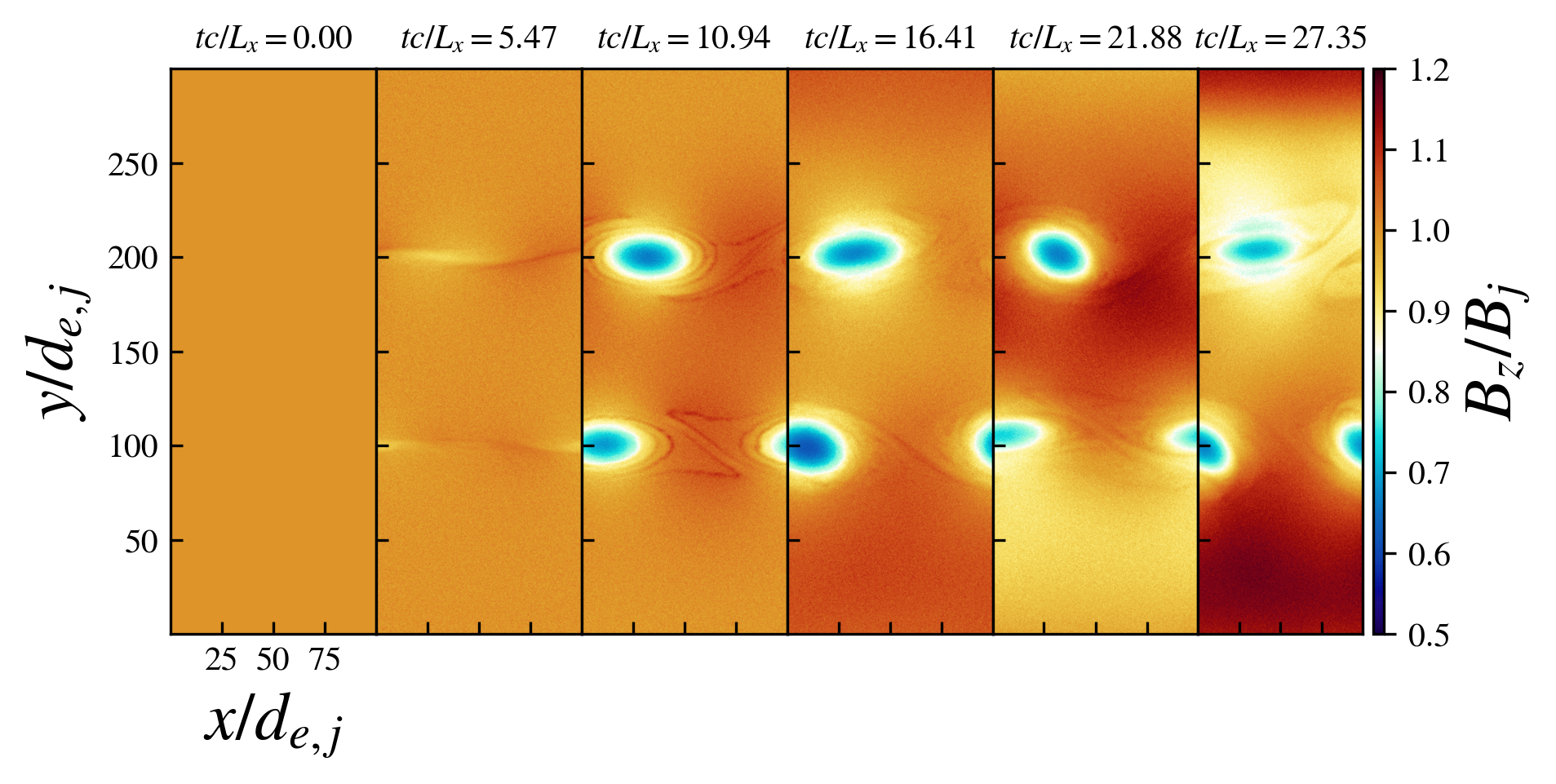}\\
    \includegraphics[width=0.45\textwidth]{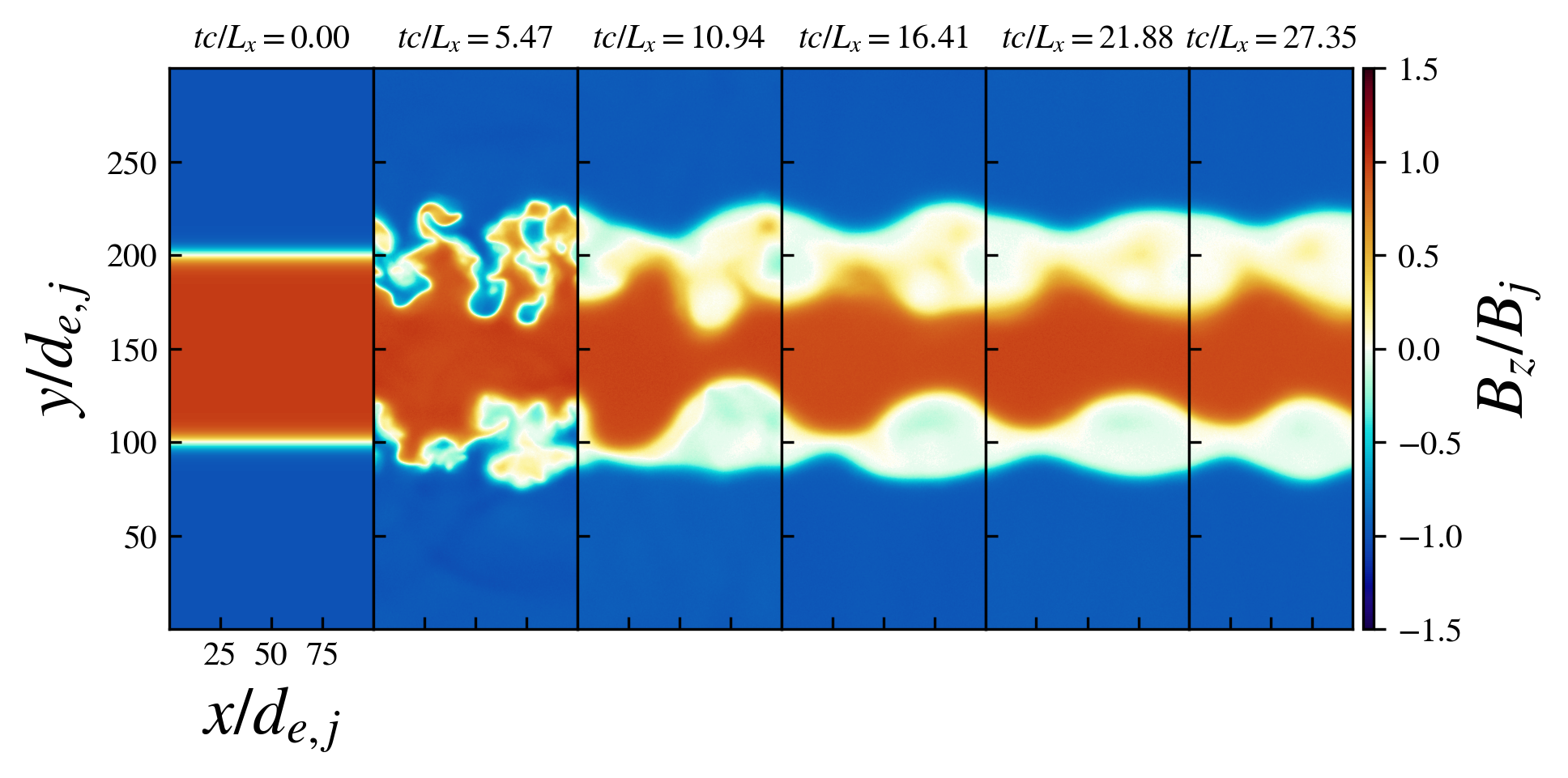} \\
    \includegraphics[width=0.45\textwidth]{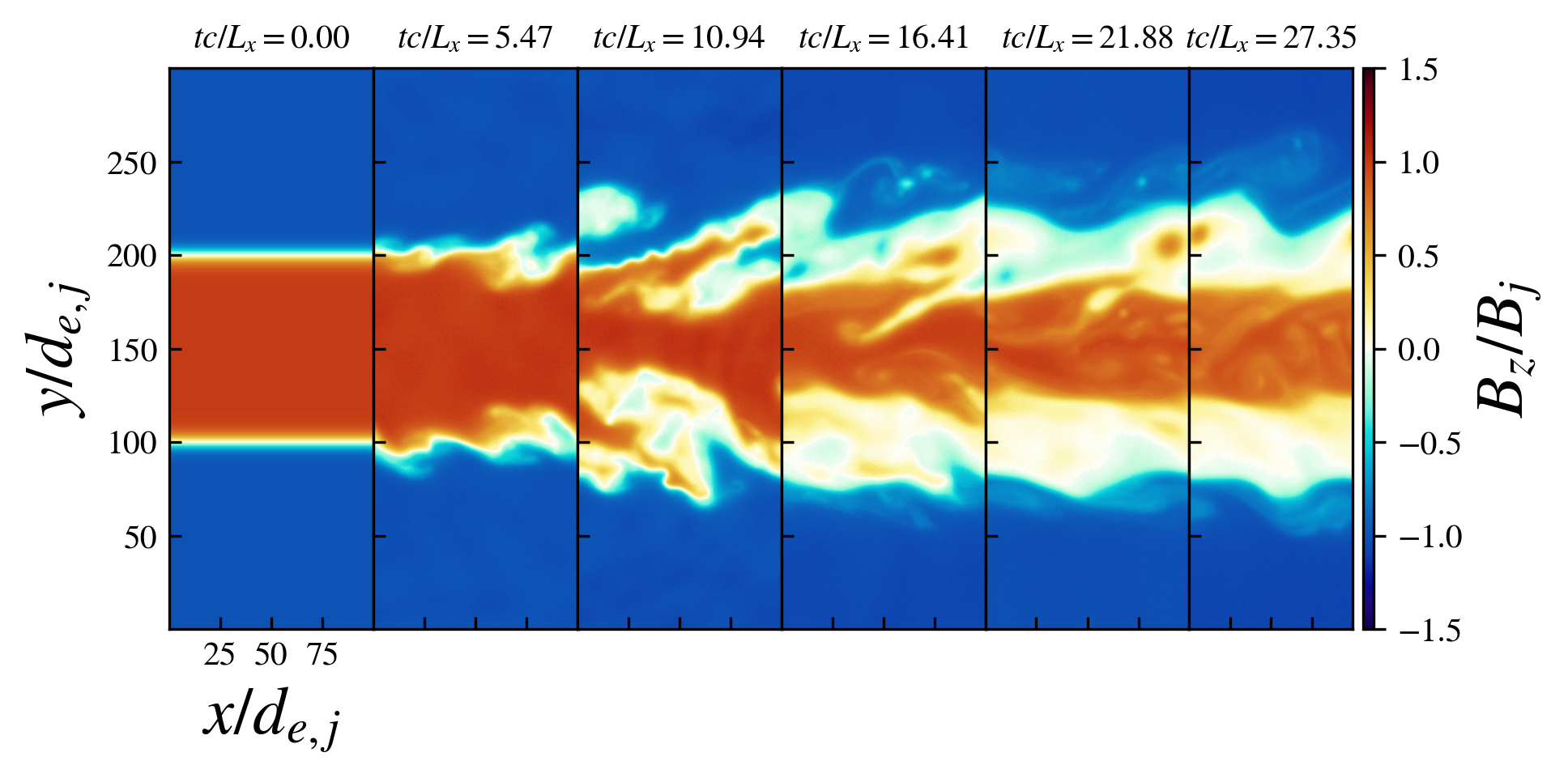} \\
    \includegraphics[width=0.35\textwidth]{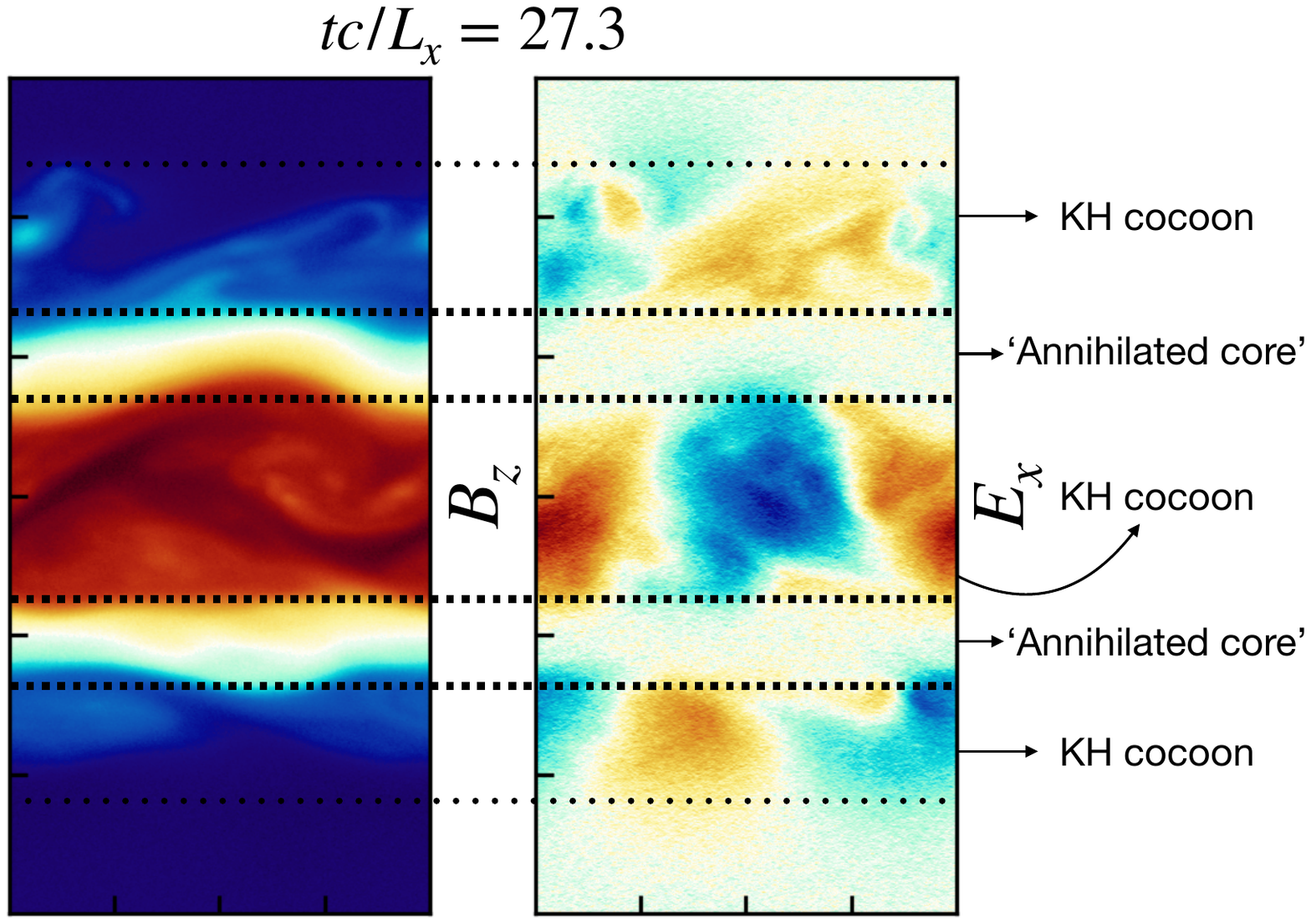}
    \caption{Snapshots of $B_z/B_j$ for simulations with velocity shear only (`VS', upper-top), magnetic shear only (`MS', upper-middle), and both shears ($u_j=0.5,B_w/B_j=-1$, upper-bottom). The lower panel shows snapshots of $B_z, E_x$ in the saturated stage ($tc/L_x=27.3$) for the $u_j=0.5,B_w/B_j=-1$ case, with the `annihilated core' and `KH cocoon' annotated.}
    \label{fig:snap_Bz}
\end{figure}

\emph{Results.}--Fig.\ref{fig:snap_Bz} shows the typical evolution of our simulations, which first follows a linear stage ($0<tc/L_x\lesssim 5-6$), then a nonlinear stage ($5-6\lesssim tc/L_x<13-20$) when the conversion of magnetic or bulk kinetic energy into internal energy happens, 
and finally a saturated stage ($tc/L_x\gtrsim13-20$). The upper-top two panels show the evolution of the two control cases: the velocity-shear-only (VS) case ($B_j/B_w = 1$), where only the KHI is excited, and the magnetic-shear-only (MS) case ($u_j=0$), where only the DKI is excited. The instabilities give rise to prominent features at the end of the linear stage: cat's-eye vortices can be clearly seen in the VS case (upper-top panel of fig.\ref{fig:snap_Bz}), while Rayleigh-Taylor-like plume features (upper-middle panel of fig.\ref{fig:snap_Bz}) arise in the MS case. The nonlinear phase begins when the $y$ displacement of the features becomes comparable to the modal wavelength in the $x$-direction. 
For the VS case this is characterized only by a slight smearing of the KH vortices, while the DK plumes mix together violently. This is the stage where most of the dissipation (if any) occurs. 
In the later saturated stage, the KH vortices persist to the end in the VS case, while in the MS case the mixing leads to a thickened shear layer where the \emph{E} and \emph{B} fields are drastically reduced. The DK plumes do not persist to the end.

When both types of shear are present, the KH and DK instabilities interact. The nonlinear stage ($tc/L_x\gtrsim 5-6$) is marked by the nonlinear interactions of the KH vortices and DK plumes, creating a turbulent shear layer (e.g. upper-bottom panel of fig.\ref{fig:snap_Bz}) at around $7-10 L_x/c$ (a small multiple of the DK growth timescale). As discussed below, the dissipation level is also modified significantly. In the saturated stage, the turbulence subsides, giving way to a shear layer consisting generally of an `annihilated core' where the $E$ and $B$ fields are suppressed, the flow is nearly stagnant, and the thermal pressure dominates. This relatively quiet layer is enshrouded by an active `KH cocoon' (lower panel of fig.\ref{fig:snap_Bz}).

\begin{figure}
    \centering
    \includegraphics[width=0.23\textwidth]{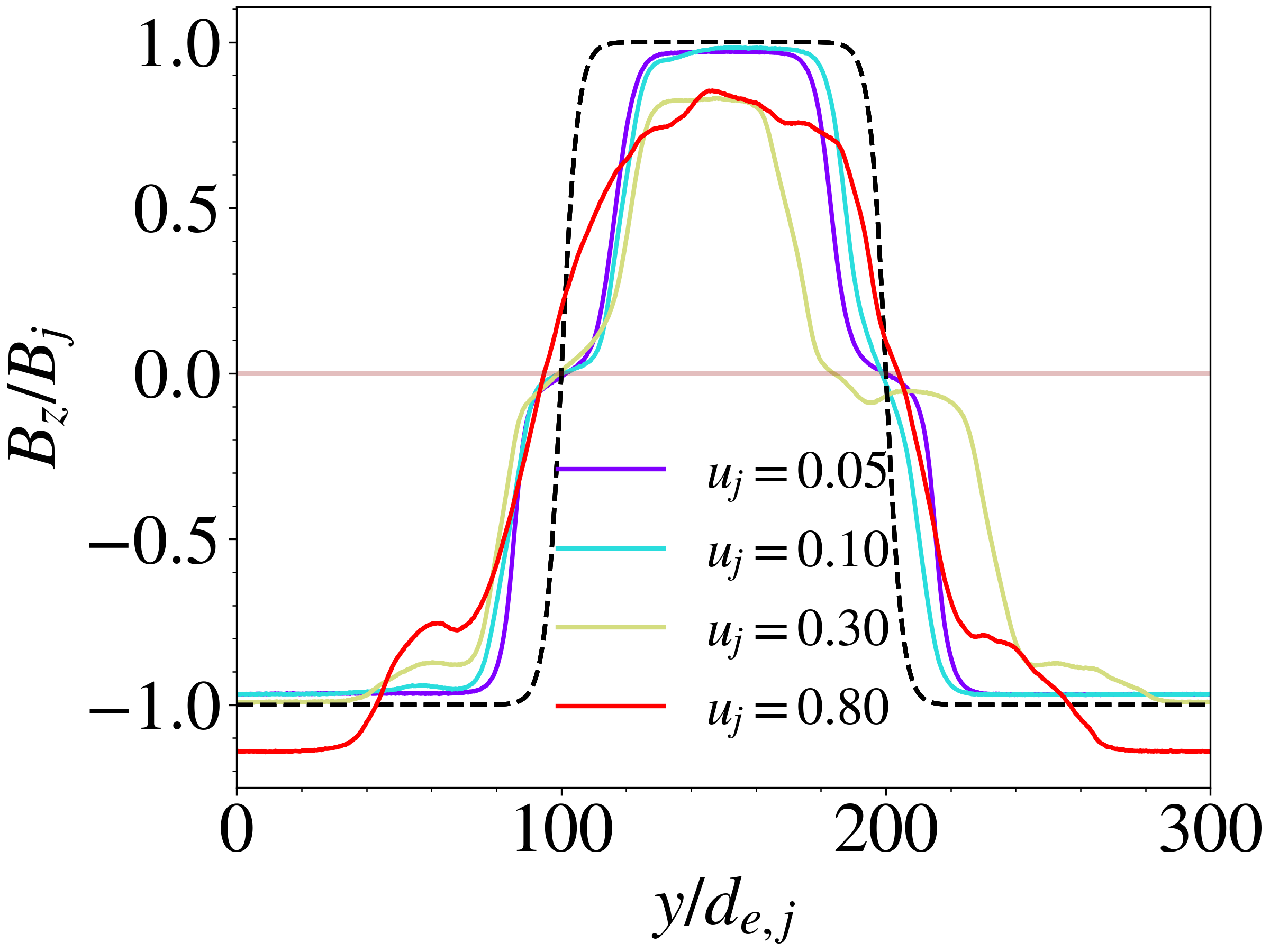}
    \includegraphics[width=0.23\textwidth]{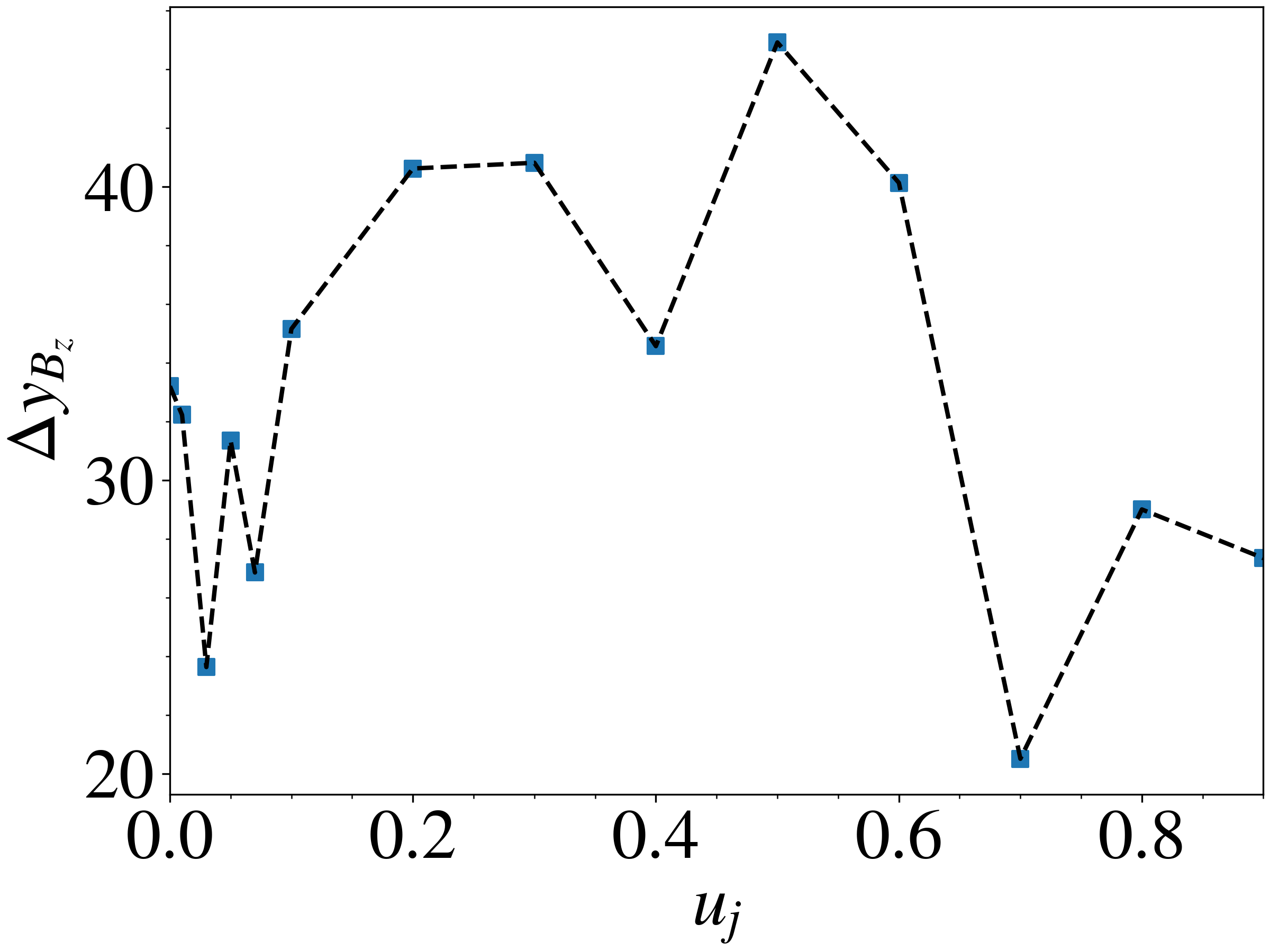} \\
    \caption{
    Left: $x$-averaged plots of $B_z/B_j$ for $u_j = 0.05,0.1,0.3,0.8$, $B_w/B_j = -1$, showing how the instabilities thicken the magnetic shear layer.
    The black dashed line is the $x$-averaged plot of $B_z/B_j$ at $t=0$. The horizontal translucent brown line indicates $B_z=0$. Right: Width of the thickened shear layer as a function of velocity shear, 
    taken at $tc/L_x=55$.}
    \label{fig:growth}
\end{figure}

We characterize the morphology of the test cases further by the width of the thickened shear layer. On the left panel of fig.~\ref{fig:growth} we plot the $x$-averaged profiles of $B_z/B_j$ at the final time $tc/L_x = 55$ for several values of~$u_j$. We observe that nonlinear interactions of the instabilities have thickened the shear layer, within which $B_z$ is reduced. We measure the width of this layer $\Delta y_{B_z}$ at $tc/L_x=55$ by the separation between $B_z/B_j = -0.5$ and $0.5$ and display it as a function of~$u_j$ in the right panel of fig.~\ref{fig:growth}. A nontrivial picture emerges. The width is reduced for a weak shear $u_j<0.1$, bounces back up for intermediate values $u_j$ ($0.2<u_j<0.6$), then drops again for larger values of~$u_j$. This is quite different from the $u_j$-dependence of the linear growth rate (not shown here for brevity), which shows no such rebound. We therefore attribute the changes in the shear width to nonlinear interplay. Namely, when the velocity shear is weak ($u_j<0.1$), DK plumes protruding across the plasma's bulk flow are quickly sheared away. As $u_j$ increases to around~0.2, KH begins to dominate and the cat's-eye vortices wrap up the layer, causing it to thicken again. When the velocity difference becomes super-magnetosonic, however, KH is once again suppressed \cite{Chow_etal-2023}, leading to a thinner saturated layer. As shown below, the width of the thickened shear layer correlates strongly with the magnetic energy dissipation.

\begin{figure}
    \centering
    \includegraphics[width=0.23\textwidth]{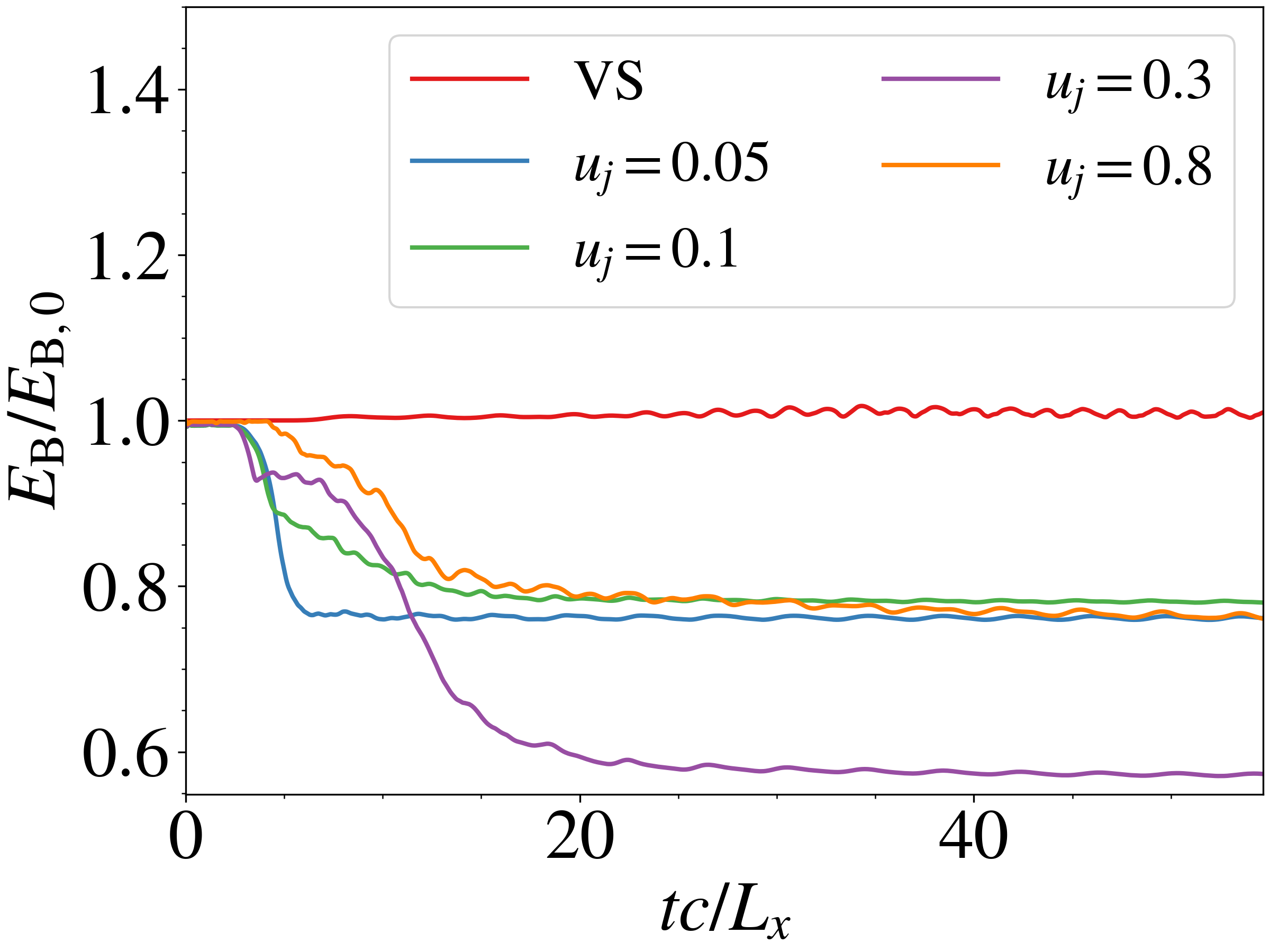}
    \includegraphics[width=0.23\textwidth]{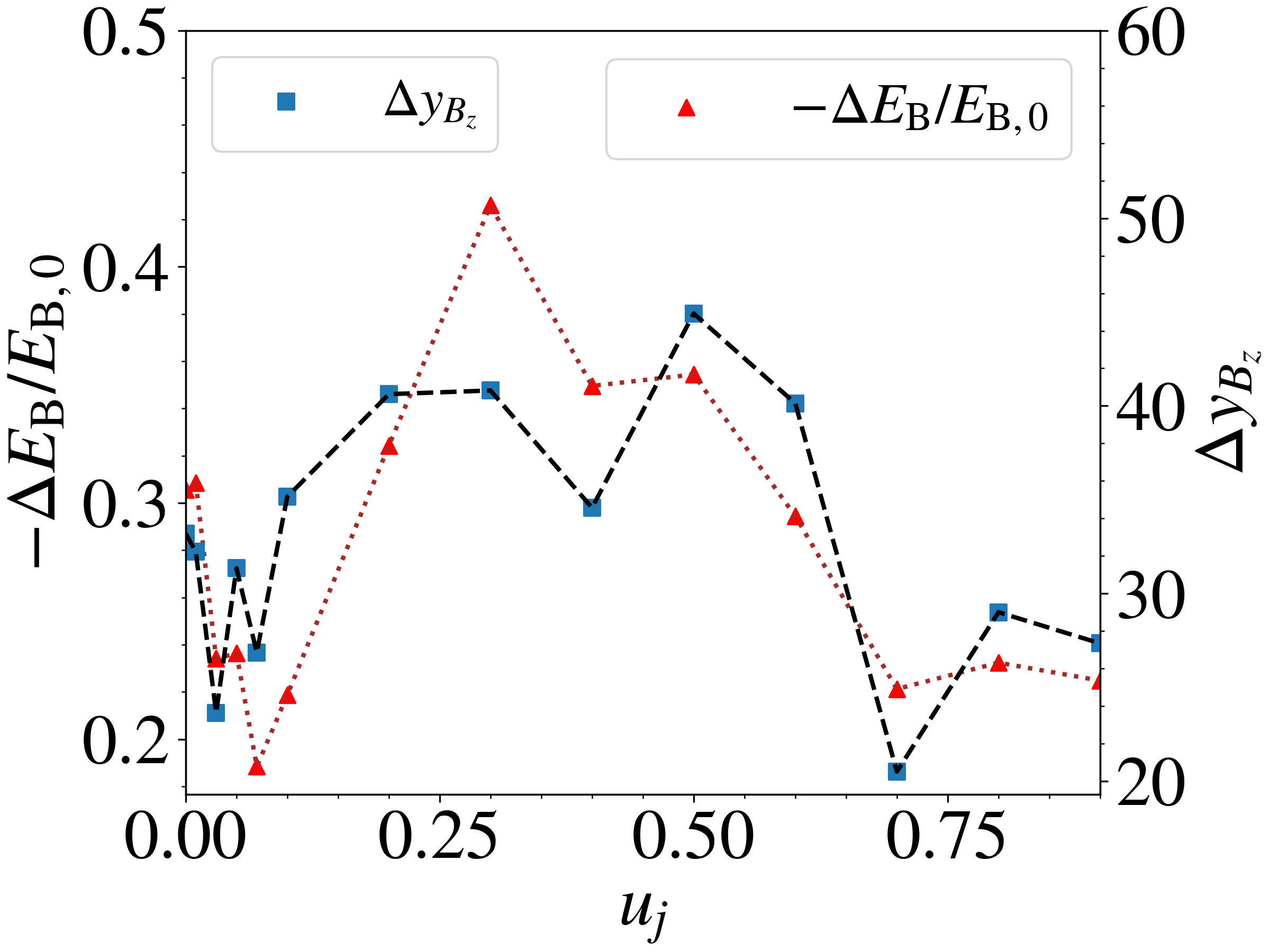}\\
    \includegraphics[width=0.23\textwidth]{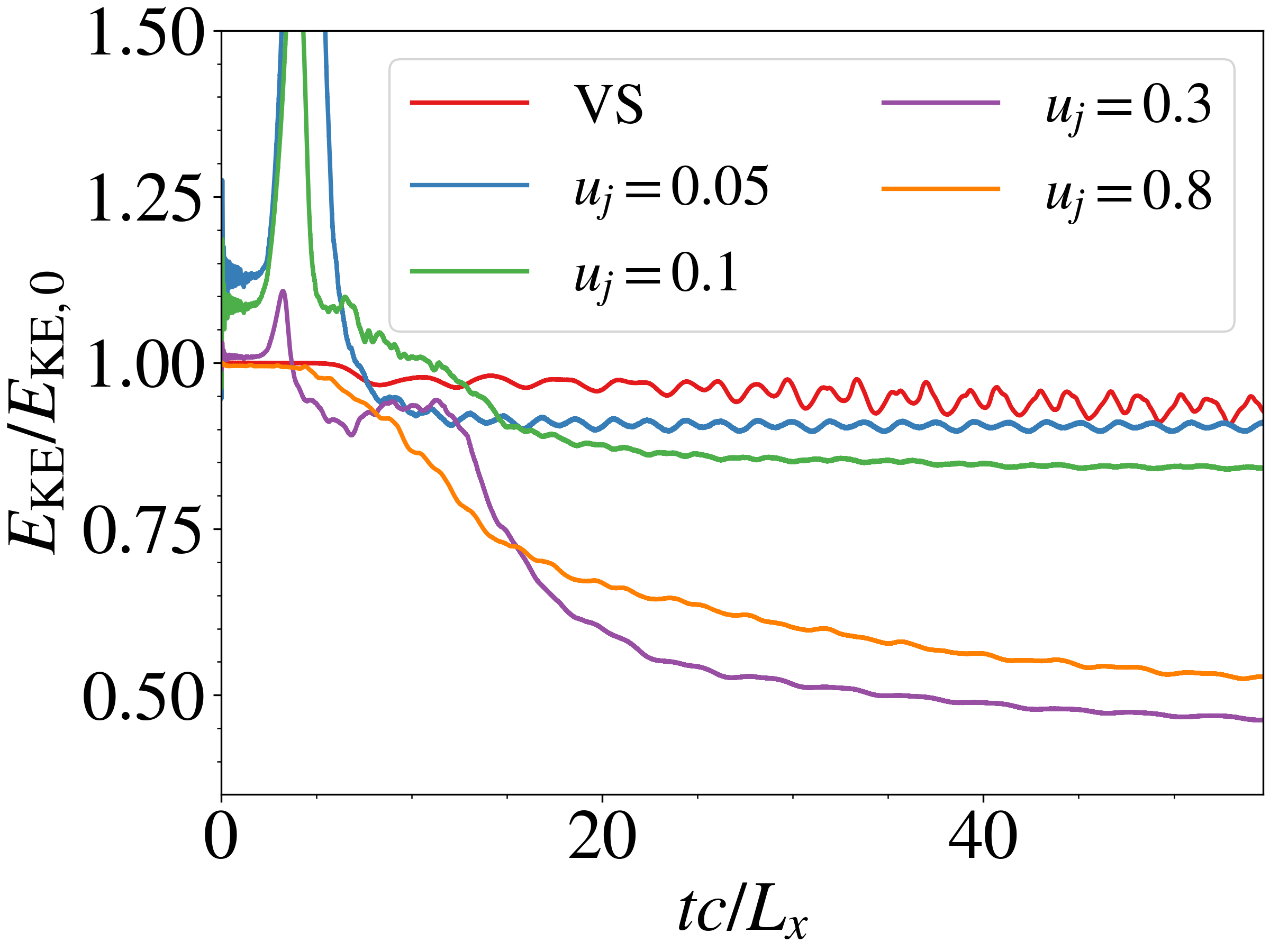}
    \includegraphics[width=0.23\textwidth]{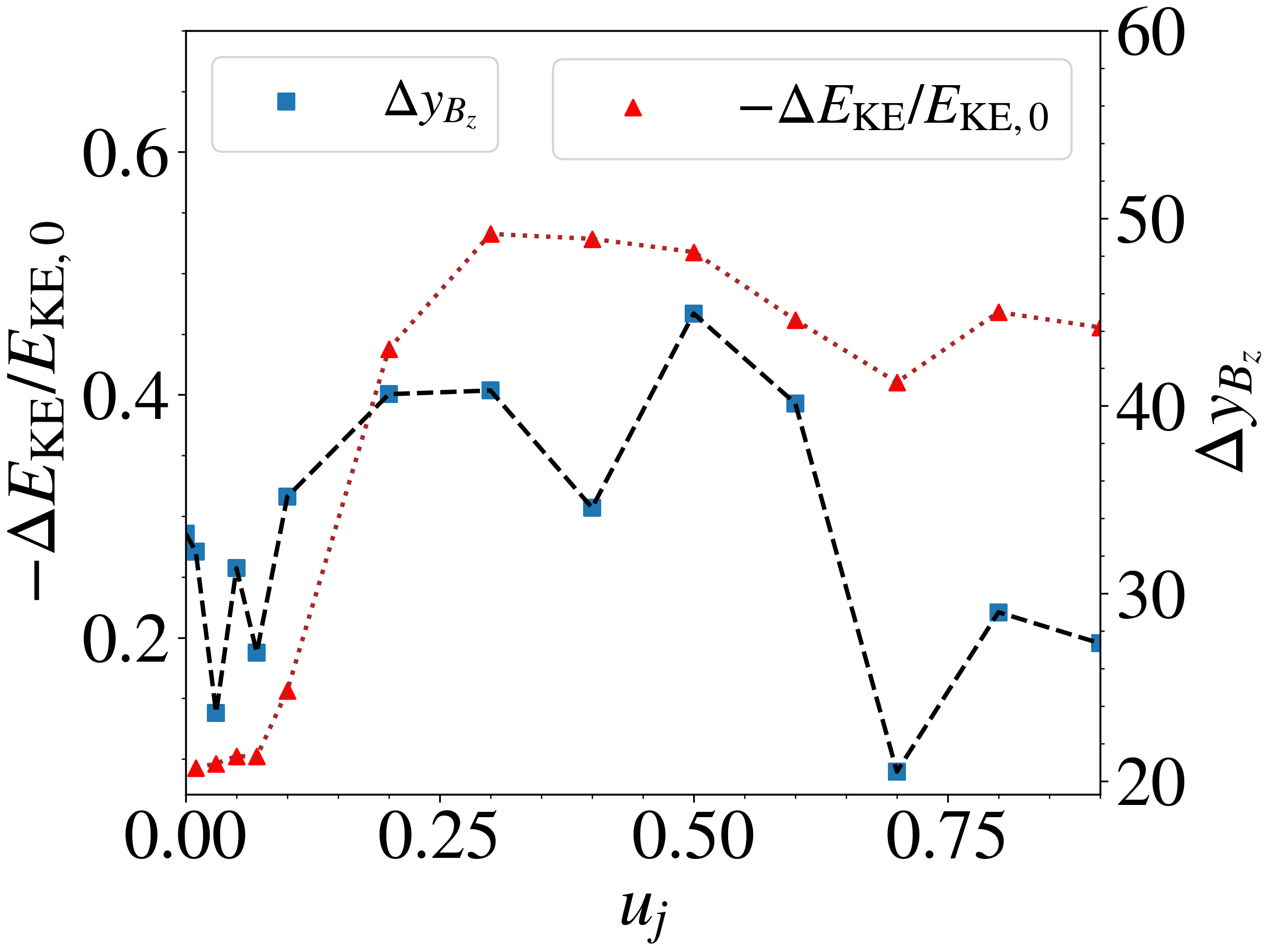}\\
    \caption{Top and bottom left: The total magnetic and bulk kinetic energies within the simulation box $E_B, E_\mathrm{KE}$, normalized by their initial values $E_{B,0}, E_\mathrm{KE,0}$, against time for selected cases ($u_j = 0.05,0.1,0.3,0.8$, $B_w/B_j = -1$ and the VS case). Top and bottom right: brown dotted lines with red triangles show the magnetic and bulk kinetic energy dissipated, measured by $-\Delta E_B/E_{B,0},-\Delta E_\mathrm{KE}/E_\mathrm{KE,0}$, as a function of velocity shear~$u_j$, taken at $tc/L_x=50$. Black dashed lines with blue squares show the width of the thickened shear layer for different~$u_j$, same as the bottom right panel of fig.\ref{fig:growth}, superimposed for comparison.}
    \label{fig:dissipate}
\end{figure}

\begin{figure}
    \centering
    \includegraphics[width=0.49\textwidth]{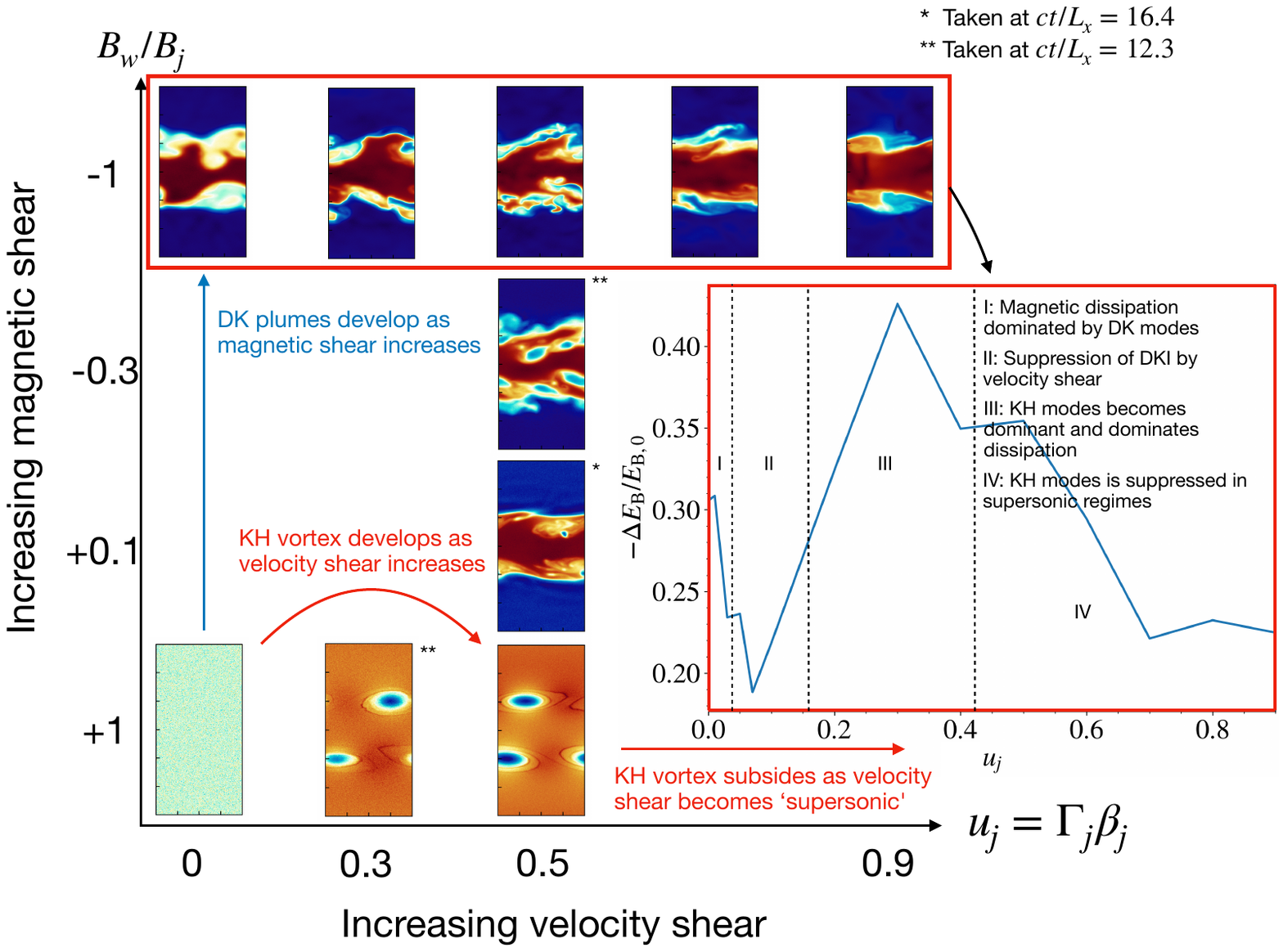}
    \caption{Schematic diagram showing how velocity and magnetic shear affect the plasma flow through the instabilities they excite, with the magnitude of the velocity shear (measured by~$u_j$, the `jet' 4-velocity) on one axis and the magnetic shear (measured by $B_w/B_j$) on the other. The snapshots are taken mostly at $tc/L_x=8.2$, with some taken at other times. The inset shows the fraction of magnetic energy dissipated as a function of $u_j$ (measured by $-\Delta E_B/E_{B,0}$, same as fig.\ref{fig:dissipate}) for the cases framed in red. The plot is divided into 4 regions, representing different regimes where a certain instability is active/suppressed.}
    \label{fig:schematic}
\end{figure}

The dissipation of magnetic and bulk kinetic energy is closely tied to the instabilities. In the top and bottom left panels of fig.\ref{fig:dissipate} we show the evolution of the total magnetic and bulk kinetic energies, $E_B$ and $E_\mathrm{KE}$, normalized by their initial values, for selected cases. Most of the dissipation occurs at around $tc/L_x\approx 10$, when the perturbations have fully grown, with $E_B$ and $E_\mathrm{KE}$ steadying out afterwards. Up to half of the initial $E_B$ and $E_\mathrm{KE}$ can be dissipated as a result of the instabilities, but as shown in the top and bottom right panels of fig.\ref{fig:dissipate}, this is very sensitive to the velocity shear~$u_j$. In fact, the dissipation of magnetic energy for different $u_j$ closely mimics the trend for the width of the thickened shear layer. The dissipation of bulk kinetic energy also shows a prominent peak at moderate velocity shears $0.2<u_j<0.6$, but unlike $E_B$, the dissipation of $E_\mathrm{KE}$ is less effective for weak velocity shears $u_j<0.1$. We note also that the case with only velocity shear (VS) has zero magnetic dissipation and close to zero (less than $5\%$) bulk-kinetic energy dissipation. While KH vortices alone do not appear to be able to dissipate magnetic energy or bulk kinetic energy effectively in this 2D configuration (at least within the timescales we explored), and even a weak velocity shear is enough to reduce the dissipation of magnetic energy with DK plumes alone, their combined effects are strong. There are two noteworthy points here: firstly, that the dissipation correlates with the thickening of the shear layer, and secondly, that the instabilities can synergistically increase the dissipation. With two shears, which could source two different instabilities, it is possible for one type of shear to limit how big the perturbations arising from the other type can grow, thus reducing dissipation; e.g. velocity shear disrupts DK plumes. At the same time, it is also possible for one type of shear to enhance the dissipative effect of perturbations arising from the other type of shear, as in the case of magnetic shear influencing KH vortices. We summarize our discussion up to this point with a schematic diagram (fig.\ref{fig:schematic}), showing how the flow morphology and dissipation change with respect to the magnitude of the two shears we investigated, within the regimes we explored.

\begin{figure}
    \centering
    \includegraphics[width=0.23\textwidth]{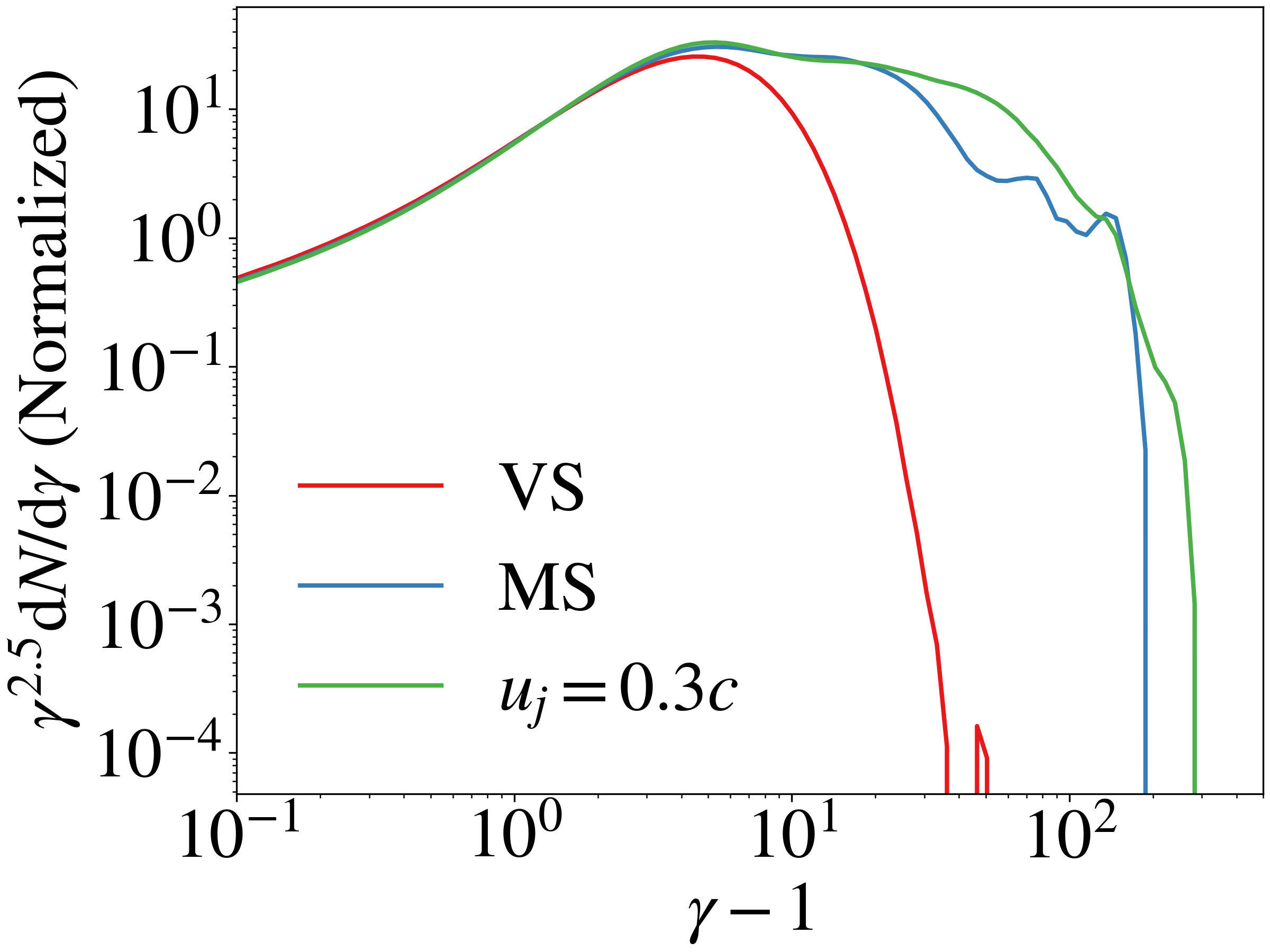}
    \includegraphics[width=0.23\textwidth]{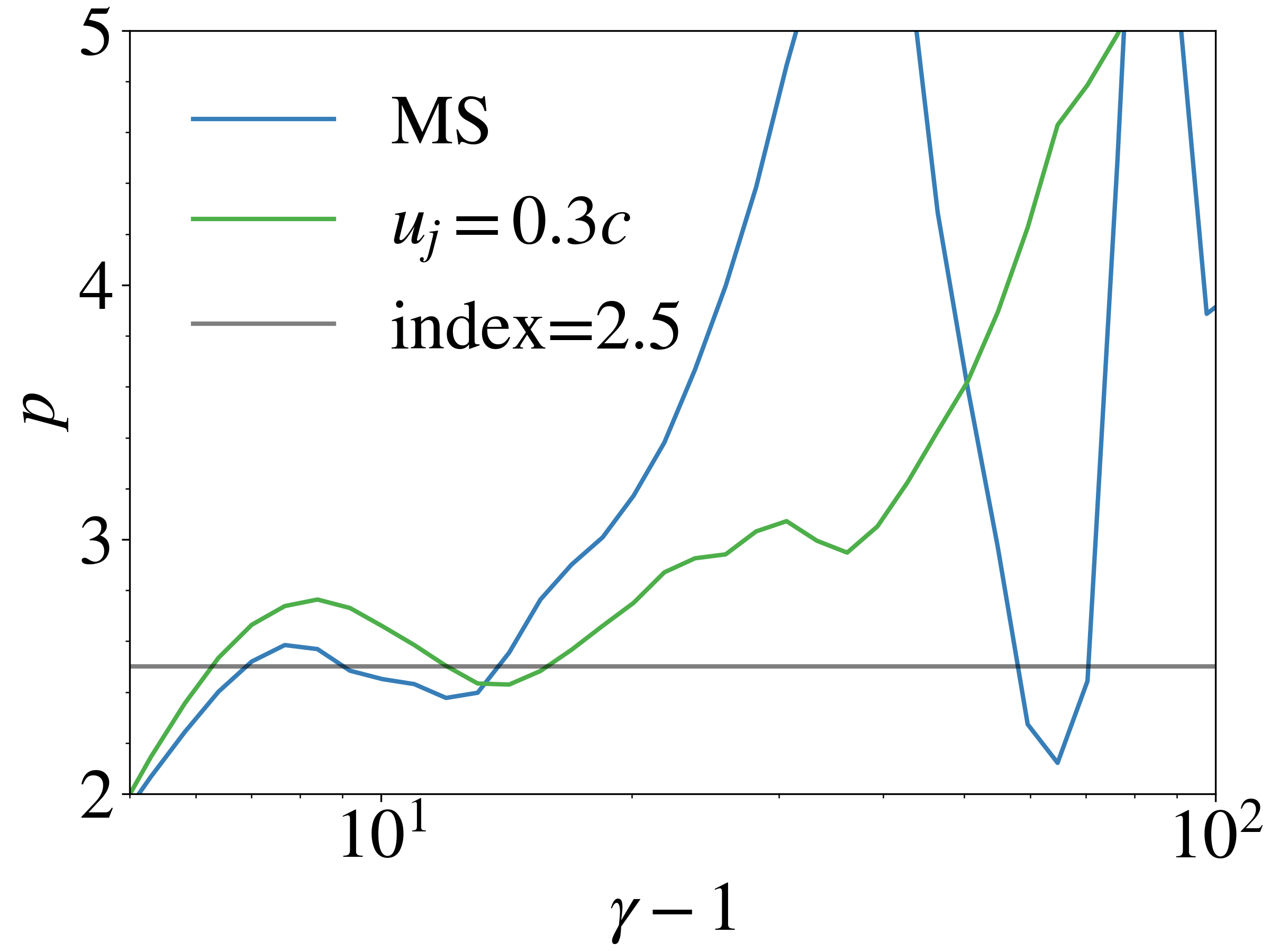}
    \caption{Left: Particle energy distributions for the VS, MS and $u_j=0.3, B_w/B_j=-1$ cases at $tc/L_c=27.3$, showing the generation of a crude nonthermal power-law like tail. Right: The power-law index $p\equiv -\mathrm{d}\log{f}/\mathrm{d}\log(\gamma - 1)$ as a function of particle energy ($\gamma - 1$) for the MS and $u_j=0.3c$ cases (the VS case is not displayed in this panel as the distribution is roughly thermal). The horizontal translucent black line corresponds to index $p=2.5$.}
    \label{fig:power_law}
\end{figure}

\begin{figure}
    \centering
    \includegraphics[width=0.23\textwidth]{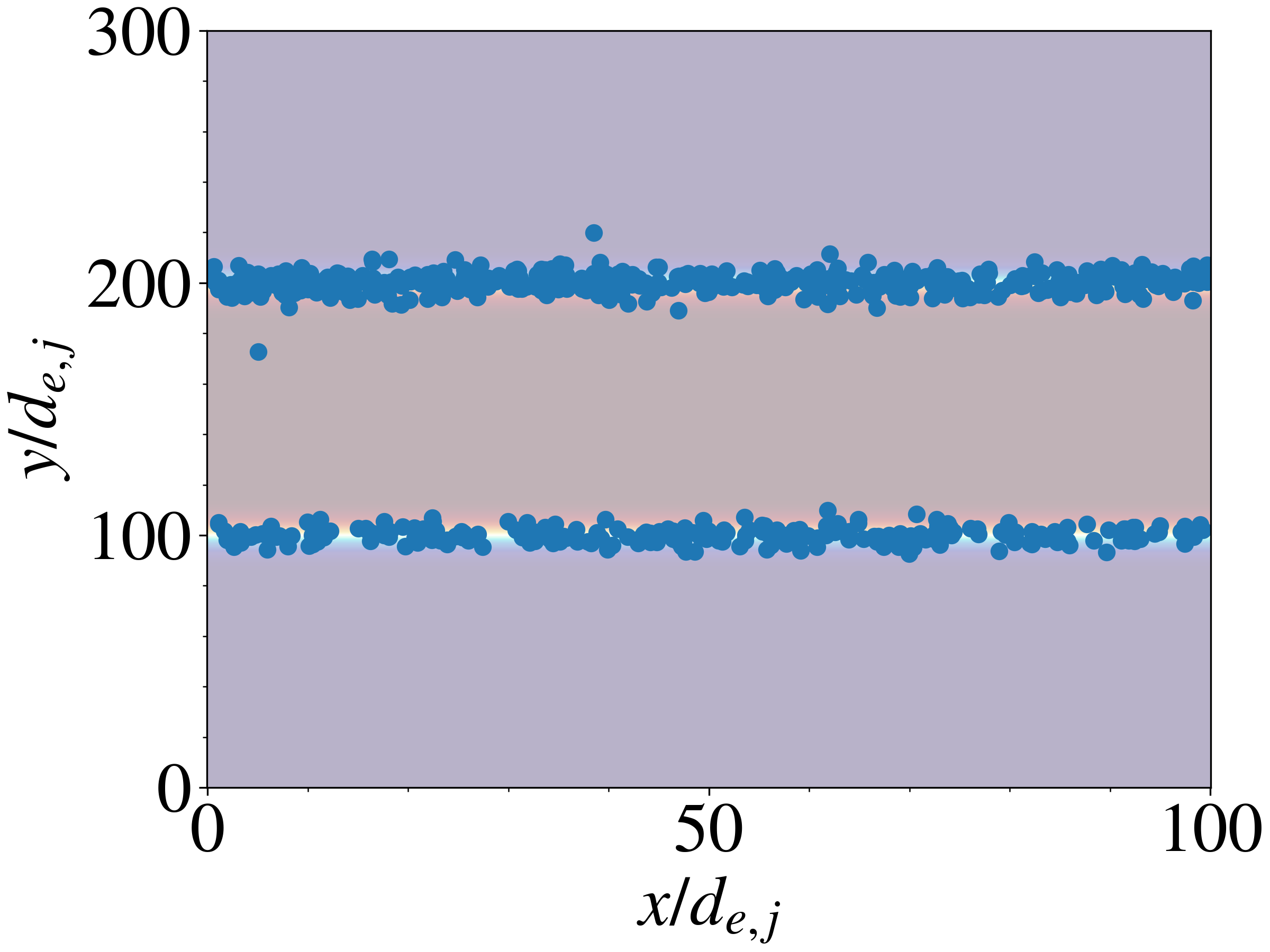}
    \includegraphics[width=0.23\textwidth]{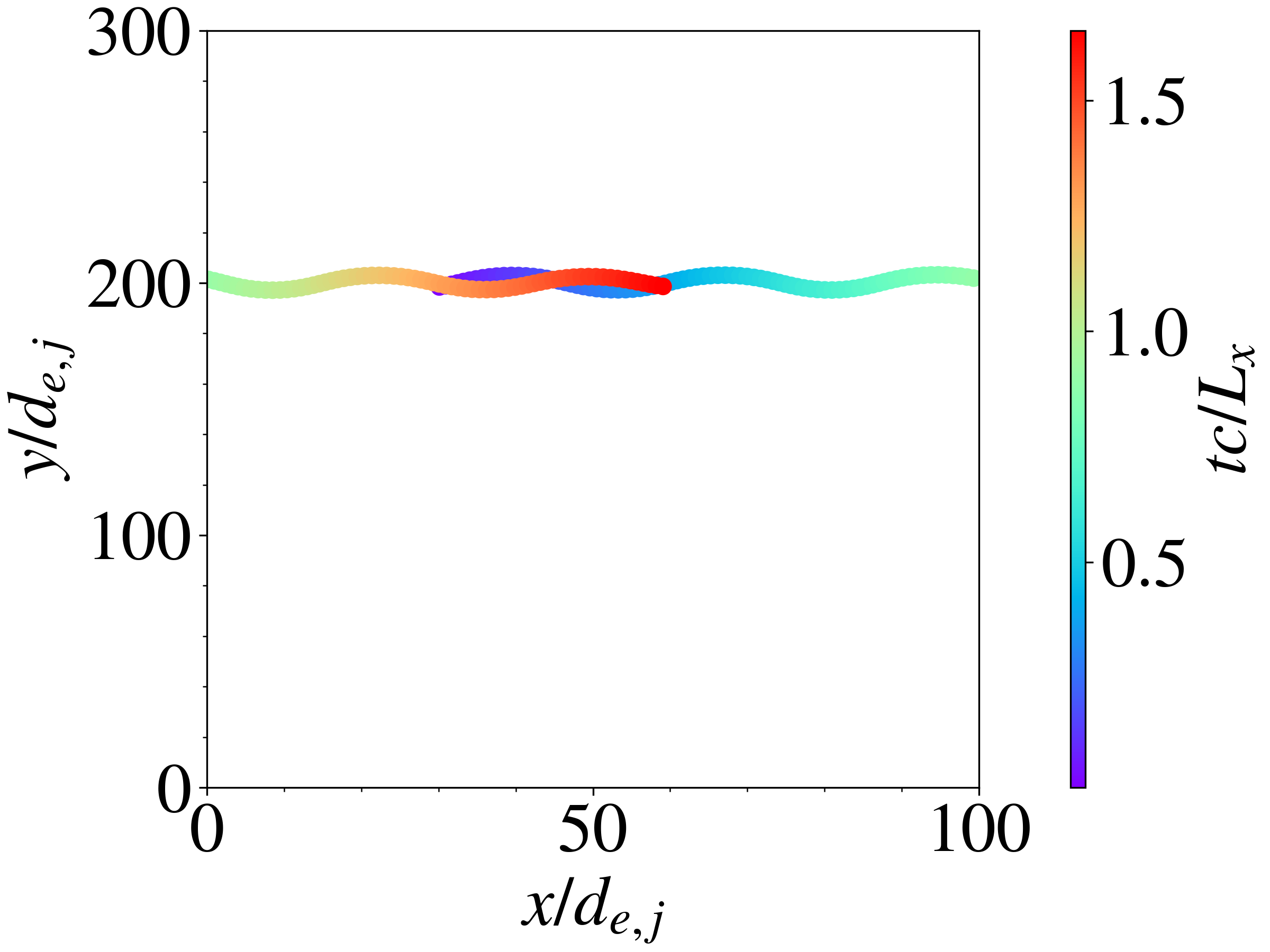} \\
    \includegraphics[width=0.23\textwidth]{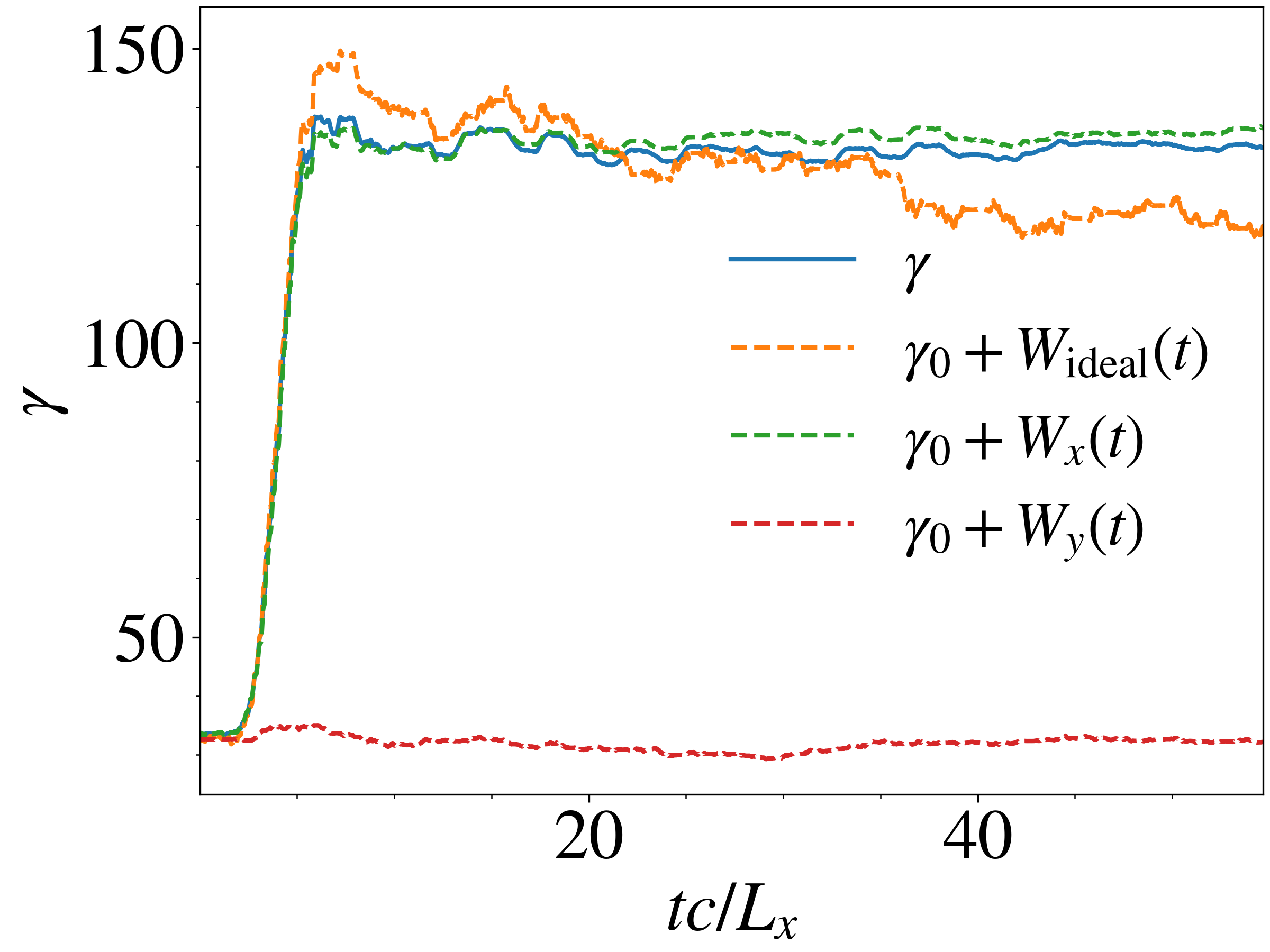}
    \includegraphics[width=0.23\textwidth]{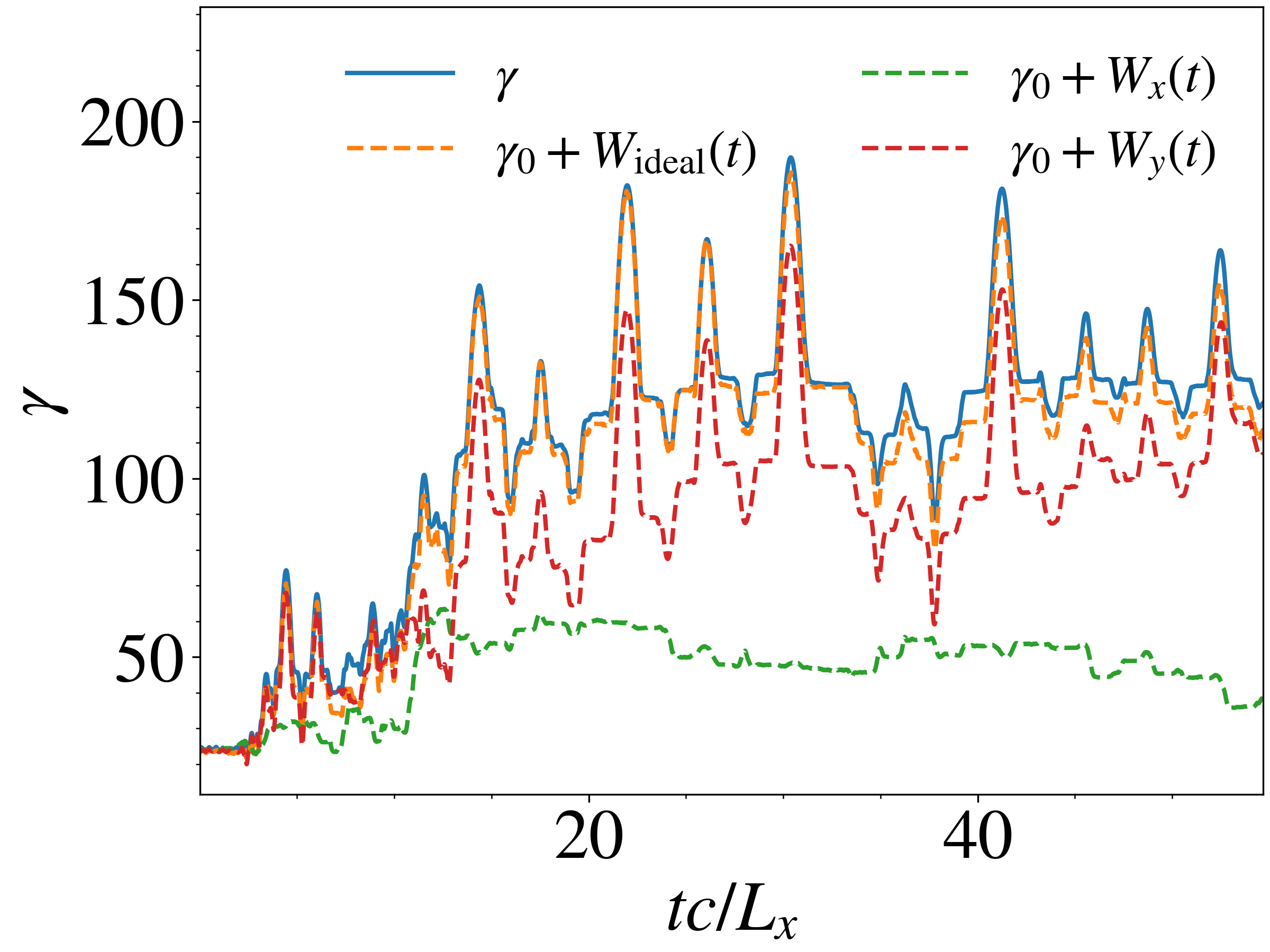}
    \caption{Top left: Initial positions of particles (blue dots) that were accelerated to $\gamma>30$, superimposed on the initial $B_z$ background for the $u_j=0.3, B_w/B_j=-1$ case. Top right: Trajectory in the first 1.6 light-crossing time of a randomly chosen accelerated particle (color-coding, from blue to red, indicates the progress of time). The particle exited the right $x$-boundary and reappeared from the left boundary during this time period.
    Bottom row: $\gamma(t)$ (solid blue line) of a randomly chosen accelerated particle, compared against the work done by various components of the $E$-field: 
    ideal-MHD, motional electric field, $W_\mathrm{ideal}(t) = -e\int_0^{t}\vb{E}_\mathrm{ideal}\cdot\vb{v}\dd{t'}/m_e c^2$ (oranged dashed line); 
    $E_x$-component, $W_x(t) = -e\int_0^{t}E_x v_x\dd{t'}/m_e c^2$ (green dashed line); 
    and $E_y$-component, $W_y(t) = -e\int_0^{t}E_y v_y\dd{t'}/m_e c^2$ (red dashed line) 
    for the MS (left) and the $u_j=0.3, B_w/B_j=-1$ (right) cases.}
    \label{fig:particle}
\end{figure}

We now delve deeper into the kinetic aspects of the instabilities and dissipation by tracking particles that have been accelerated. We first point out the generation of nonthermal power-law tails with a spectral index of about~2.5 for the MS and $(u_j=0.3, B_w/B_j=-1)$ cases in fig.\ref{fig:power_law}, suggestive of a scale-invariant dissipative process. We then demonstrate, in fig.\ref{fig:particle}, how particles are accelerated by different components of the electric field as they follow Speiser-like orbits close to the shear interface. In the top left panel of fig.\ref{fig:particle} we display the initial positions of the particles that underwent substantial acceleration, defined by acquiring a Lorentz factor above~30 by $tc/L_x=27.4$, i.e., much higher than the mean $\bar{\gamma} \simeq 2.4$ corresponding to a thermal distribution with temperature $\theta=1$. These particles are clustered initially at the shear interface, implying that they are likely accelerated there. In the top right panel of fig.\ref{fig:particle} we show the first light-crossing-time segment of the trajectory of a randomly selected accelerated particle that has exceeded $\gamma=100$ at some point in time. The trajectory's curvature is reversed every time the particle crosses the interface, effectively surfing the shear layer in the $x$-direction at almost the speed of light. In the bottom panels of fig.\ref{fig:particle} we track the Lorentz factors of randomly selected accelerated particles for the MS case (bottom left) and the $(u_j=0.3, B_w/B_j=-1)$ case (bottom right), and compare them with the work done by various components of the electric field. We first remark that particle acceleration is done predominantly by the ideal-MHD, motional electric field in both cases. In the MS case,  acceleration is dominated by the work done by~$E_x$, the perturbed electric field generated entirely by~DKI. In the mixed-shear $(u_j=0.3, B_w/B_j=-1)$ case, the situation is more complex as we observe concurrent acceleration by both $E_x$ and~$E_y$. We also observe spikes due to the work done by the background motional $E_y$ field, which is potential, with the particle's energy being roughly the same at the start and end of each spike. Due to the relative difficulty of separating the work done by the perturbed $E_y$ from the background, in the following we will focus mainly on~$E_x$, relegating a more thorough investigation of particle acceleration to a future study. 

\begin{figure}
    \centering
    \includegraphics[width=0.23\textwidth]{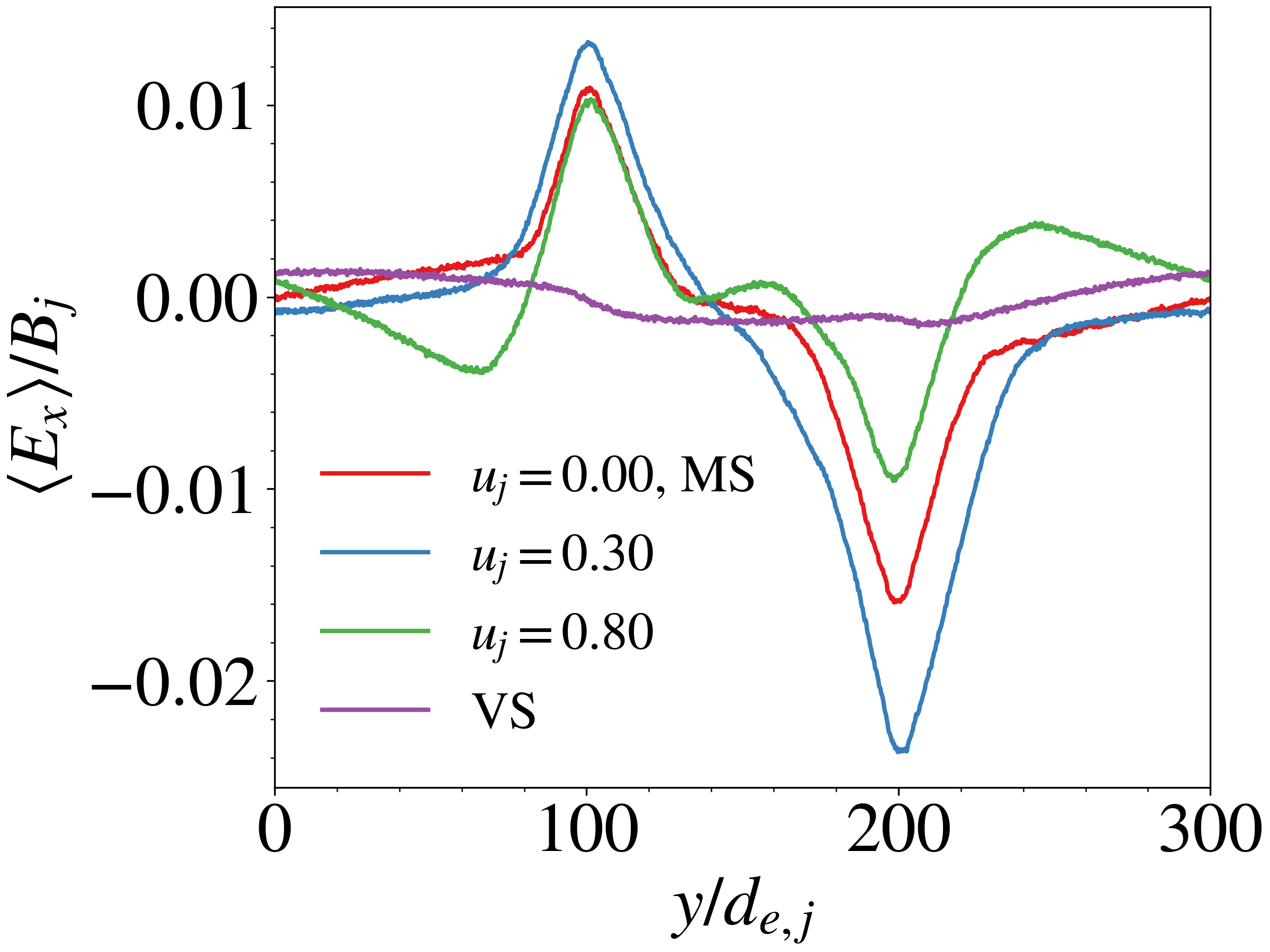} 
    \includegraphics[width=0.23\textwidth]{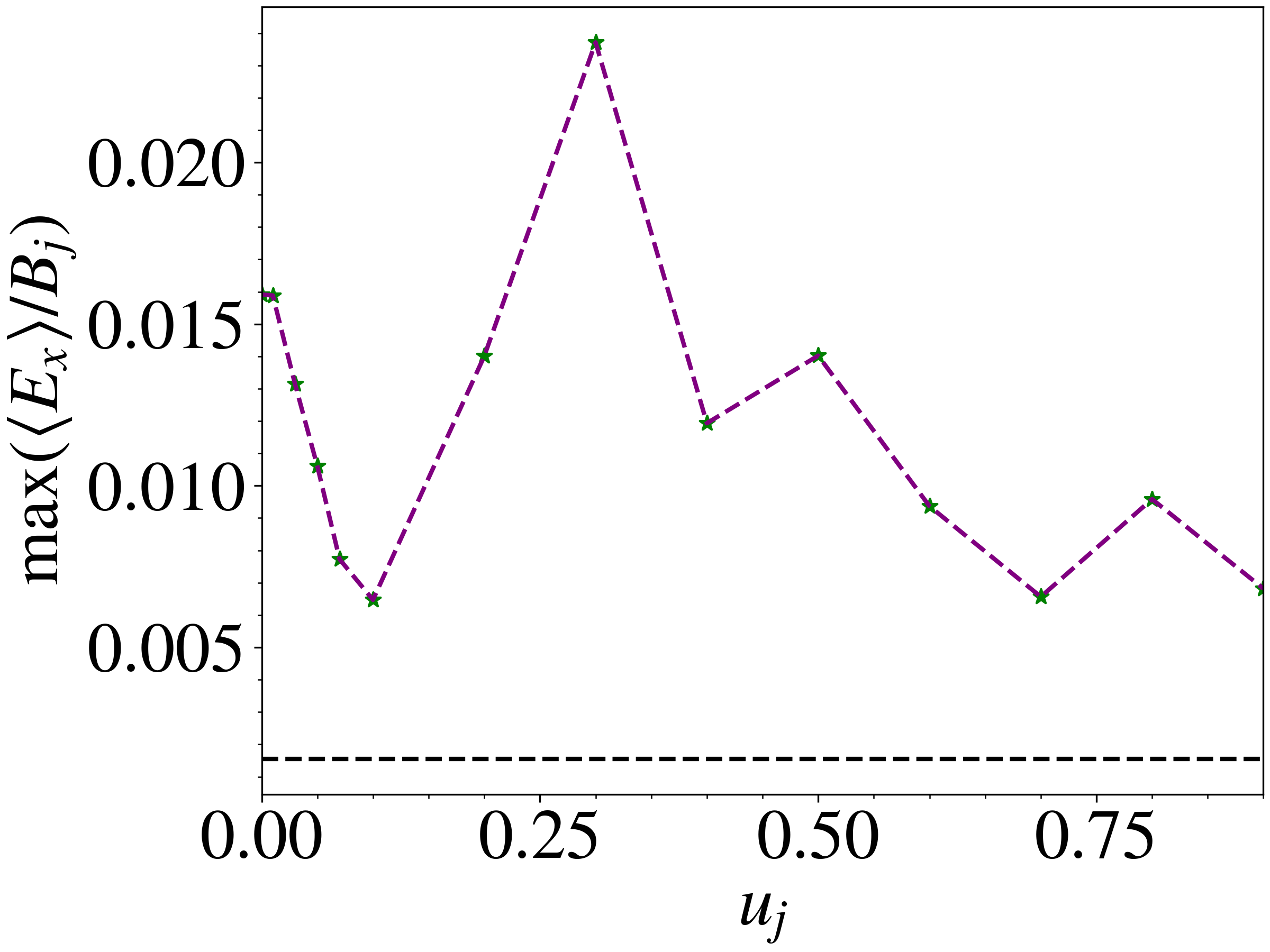} 
    \caption{Left: Time- and $x$-averaged profiles of $E_x(y)$ for selected cases ($u_j = 0.3, 0.8, B_w/B_j=-1$ and the MS, VS cases). The time-averaging is carried out from $tc/L_x=0$ to $13$, when most of the dissipation occurred. Right: The peak magnitude of the time- and $x$-averaged profiles of $E_x(y)$ between $190<y/d_{e,j}<210$, against~$u_j$. The black dashed line indicates such value for the VS case.}
    \label{fig:time_average}
\end{figure}

Obviously, dissipation depends on the magnitude of the accelerating electric field. The crucial question is what were the average $E_x$-values that the particles experienced over the course of their trajectories? On the left panel of fig.\ref{fig:time_average} we show the $x$-averaged and time-averaged profiles of $E_x(y)$ for selected cases. We use this as a proxy for the average $E_x$ experienced by the particles because most of the accelerated particles were performing Speiser-like orbits very close to the shear layer. The time average is taken from $tc/L_x=0$ to~$13$, the period for which most of the dissipation occurred. The averaged $E_x(y)$ profiles generally peak at the two shear interfaces. We note that the VS case, where only the KH instability is operational, has a near-zero averaged $E_x$ profile, despite having a significantly thickened shear layer due to the cat's-eye vortices. This is due to the periodic nature of the cat's-eye vortices, i.e. a particle surfing in the $x$-direction close to the shear interface sees periodic rises and drops in $E_x$ of the same magnitude but opposite sign, resulting in an almost perfect cancellation of $E_x$ along its path, and hence almost zero net $x$ acceleration. This is consistent with the observed weak overall dissipation in the VS case. In the right panel of fig.\ref{fig:time_average} we display the peak magnitude of the time- and $x$-averaged $E_x$ in the interval $190<y/d_{e,j}<210$ as a function of velocity shear~$u_j$. The peak magnitude for the VS case is displayed by a black dashed line. This plot shows that simultaneous operation of the KH and DK instabilities, as opposed to just~KH, is needed to generate a substantial averaged~$E_x$. Our observations suggest that the presence of a magnetic shear, which generated an additional instability~(DK), broke the periodicity of the KH vortices and facilitated a net average~$E_x$, which led to a substantial increase in particle acceleration and dissipation. 
The MS case, on the other hand, had plenty of dissipation with just the DKI alone. It turns out, as described by \cite{Zenitani_Hoshino-2007}, that periodicity-breaking of DK modes is achieved naturally as the DK plumes align $E_x$ of the same polarity in the nonlinear stage, while the reversing magnetic field confines the particles close to the shear layer. 
The bottom line is that the instability-induced E-field, which has the potential to accelerate particles, requires some periodicity-breaking mechanism to turn that potential into a net accelerating force. This can either be due to nonlinear modal interaction arising from two or more instabilities, as in the ($u_j=0.3, B_w/B_j=-1$) case, or to the natural alignment of perturbed electric field with the same polarity, as in the MS case. The result is that a particle `sees' a net perturbed electric field that is coherent over the course of its trajectory, leading to its acceleration. We see in the right panel of fig.\ref{fig:time_average} that the instability-induced averaged $E_x$ is strongest for intermediate values of $u_j$ (0.2-0.5); this suggests that KHI and DKI can work together to enhance the accelerating E-field. 
As discussed above, magnetic dissipation is also greatest in this range of~$u_j$; there is thus a strong correlation between $\langle E_x\rangle$ and magnetic dissipation.

\emph{Conclusion.}--The nonlinear interaction of the KHI and DKI generated qualitatively new structures with very different dissipative behaviors compared to when the shears are taken in isolation. We find that DKI can effectively disrupt the cat's-eye vortices generated by KHI, creating a turbulent shear layer on the DK timescale. This interplay leads to a significant enhancement of dissipation over the velocity-shear-only case.  In addition, we find a special, relatively narrow range in velocity shear where the joint DK-KHI is particularly active, resulting in even stronger dissipation. Finally, we observe efficient nonthermal particle acceleration caused by the alignment of the instability-driven electric fields with Speiser-like motion of particles close to the shear interface. This study highlights the sensitivity of flow structures, dissipation, and particle acceleration to multiple simultaneously operating instabilities, thus providing a strong motivation for further studies of their nonlinear interaction at the kinetic level. Such studies will help elucidate the nature of dissipation and particle acceleration, and may explain the observed limb-brightening of emission in relativistic jets from astrophysical supermassive black holes.

\emph{Acknowledgements.}--We thank Lorenzo Sironi and Vladimir Zhdankin for useful comments and suggestions. This work is supported by NASA ATP Grant 80NSSC22K0828 and ACCESS computing grant PHY140041.

\appendix 

\section{Details of the jet-wind model setup} \label{app:setup}

We describe the details of our setup, which is based on the one described in~\cite{Rowan-2019}. 
Note that due to velocity and magnetic shears, there will be current and charge excesses at the shear interfaces, which imply that the electron and positron densities and bulk velocities will be different close to the interfaces. Therefore, when we mention `jet' and `wind' quantities, we are referring to their values far from the interfaces, in the respective `jet' and `wind' regions, where the current and charge excess are zero. There, the bulk speeds and the rest-frame jet/wind electron and positron densities are equal: $\beta_{e,j}=\beta_{p,j}, \beta_{e,w}=\beta_{p,w}$ and $\tilde{n}_{e,j} = \tilde{n}_{p,j}, \tilde{n}_{e,w} = \tilde{n}_{p,w}$. 

Assuming equal jet rest-frame electron and positron temperatures, $\theta_{e,j} = \theta_{p,j}\equiv\theta_j$, 
the rest-frame mean particle speed $\bar{v}_j$ in the jet (and the mean Lorentz factor~$\bar{\gamma}_j$) for a given jet temperature $\theta_j$ can be determined from the local Maxwell-J\"uttner distribution. The adiabatic index $\Gamma_\mathrm{ad}(\theta_j)$
of the electrons and positrons can be found using the fitting formula given by eqn.14 in \cite{Service-1986}. 
Together,  $\bar{\gamma}_j$ and the rest-frame jet electron (and positron) particle density $\tilde{n}_{e,j} = \tilde{n}_{p,j} = \tilde{n}_{e,j}$ define the main normalizing length-scale in our simulations --- the electron inertial length $d_{e,j} \equiv c/\omega_{pe,j}$, where $\omega_{pe,j} \equiv [4\pi\tilde{n}_{e,j}e^2/\bar{\gamma}_j m_e]^{1/2}$ is the comoving electron plasma frequency.

Next, the comoving jet enthalpy density, which enters the expression for magnetization~$\sigma_j$, is given by $w_j = w_{e,j} + w_{p,j} = 2(1+\Gamma_{\mathrm{ad},j}/(\Gamma_{\mathrm{ad},j}-1)\theta_j)\tilde{n}_{e,j}m_e c^2$.  

The plasma bulk velocity $\beta_x\qty(y)$ and the magnetic field are initialized with the following profiles:
\begin{gather}
    \beta_x\qty(y) = \beta_j\qty[1 - \tanh(\frac{y-y_1}{\Delta}) + \tanh(\frac{y-y_2}{\Delta})], \\
    B_z\qty(y) = B_j\biggl[\frac{B_w}{B_j} + \frac{1}{2}\qty(1 - \frac{B_w}{B_j})\biggl[\tanh(\frac{y-y_1}{\Delta}) \nonumber\\ - \tanh(\frac{y-y_2}{\Delta})\biggr]\biggr],
\end{gather}
where $\beta_j = u_j/(1 + u_j^2)^{1/2}$ is the velocity of the jet (normalized by~$c$) corresponding to the specified normalized 4-velocity~$u_j$, $y_1,y_2$ are the $y$-locations of the shear interfaces, and $\Delta$ is the half-width of the shear (and also current) layer at these interfaces. 
%
The initial lab-frame electric field, current density, and charge density are given by $\vb{E}=-\vb{v}\cross\vb{B}/c= \beta_x B_z\vu{y}$, $\vb{J} = c\dv*{B_z}{y}/4\pi\vu{x}$, and $\rho_e = (\dv*{E_y}{y})/4\pi$, as dictated by the ideal-MHD Ohm's, Ampere's law, and Gauss's law, respectively. 

The $y$-profiles of the lab-frame electron and positron number densities $n_e, n_p$ and of the bulk velocities $\beta_{e,x}, \beta_{p,x}$ are constrained to satisfy the self-consistency conditions
\begin{gather}
    \rho_e = \qty(n_p - n_e) e, \label{eqn:rho_e}\\
    J_x = \qty(n_p \beta_{p,x} c - n_e \beta_{e, x}) e c, \\
    \beta_x = \frac{n_p\beta_{p,x} + n_e\beta_{e,x}}{n_p + n_e}. \label{eqn:mean_vel}
\end{gather}
To proceed, we recast $n_p, n_e$ as
\begin{equation}
    n_p = n_0\qty(1-\delta_n), \quad n_e = n_0\qty(1+\delta_n), \label{eqn:density_recast}
\end{equation}
where $n_0 = \Gamma(y)\tilde{n}_0$ can be regarded as some background lab-frame density profile. The bulk Lorentz-factor profile is given by $\Gamma(y) = 1/(1 - \beta_x^2)^{1/2}$, and the rest-frame background density $\tilde{n}_0$ by
\begin{gather}
    \tilde{n}_0\qty(y) = \tilde{n}_{e,j}\biggl[\frac{\tilde{n}_{e,w}}{\tilde{n}_{e,j}} + \frac{1}{2}\qty(1 - \frac{\tilde{n}_{e,w}}{\tilde{n}_{e,j}})\biggl[\tanh(\frac{y-y_1}{\Delta}) \nonumber\\ - \tanh(\frac{y-y_2}{\Delta})\biggr]\biggr].
\end{gather}
In this study, for simplicity we take the rest-frame jet and wind density to be the same, $\tilde{n}_{e,w}/\tilde{n}_{e,j} = 1$. Substituting eqn.\ref{eqn:density_recast} into eqn.\ref{eqn:rho_e}-\ref{eqn:mean_vel} and solving, we then have
\begin{gather}
    \delta_n = -\frac{\rho_e}{2n_0 e}, \\
    \beta_{e,x} = \frac{2n_0 e\beta_x - J_x/c}{2n_0 (1 + \delta_n)e}, \quad \beta_{p,x} = \frac{2n_0 e\beta_x + J_x/c}{2n_0 (1 - \delta_n)e}.
\end{gather}
Finally, to determine the electron and positron temperature profiles $\theta_{e,p}(y)$, we employ pressure balance:
\begin{equation}
    \tilde{n}_p m_e c^2\theta_p + \tilde{n}_e m_e c^2\theta_e + \frac{B_z^2}{8\pi\Gamma^2} = \mathrm{const.} \label{eqn:pressure_balance}
\end{equation}
Assuming $\theta_p(y) = \theta_e(y)$ initially, these temperatures can be determined from eqn.\ref{eqn:pressure_balance}. 
This completes the PIC setup of the jet-wind model.

\section{Definition of the plasma microscales} \label{app:microscales}

The key plasma length-scales describing our system are the initial jet electron inertial length 
$d_{e,j} \equiv c/\omega_{pe,j}$, where $\omega_{pe,j} \equiv [4\pi \tilde{n}_{ej} e^2/\bar{\gamma}_jm_e]^{1/2}$ 
is the plasma frequency ($\tilde{n}_{e,j}$ is the rest-frame jet electron number density and $\bar{\gamma}_j$ is the mean Lorentz factor of the jet rest-frame MJ distribution), 
the jet electron Debye length $\lambda_{\mathrm{D}e} \equiv (T_j/4\pi n_{e,j} e^2)^{1/2}\sim\theta_j^{1/2} d_{e,j}$, 
and the characteristic jet electron gyroradius $\rho_j \equiv \bar{\gamma}_j m_e \bar{v}_j c/e B_j\sim (\theta_j/\sigma_j)^{1/2} d_{e,j}$ (where $\bar{v}_j$ is the mean particle velocity). For relativistically warm ($\theta\sim 1$), moderately magnetized ($\sigma\sim 1$) plasma the three plasma length-scales ($d_{e,j},\lambda_{\mathrm{D}e},\rho_j$) are roughly equal.

\section{Decomposing stress tensor into thermal and bulk components} \label{app:decompose}

We describe the procedure for decomposing the stress tensor into thermal and bulk-flow components, needed to monitor the bulk-kinetic energy content. We thank Vladimir Zhdankin (private communication) for offering insights on this procedure.

The procedure is as follows:
\begin{enumerate}
    \item[1.] For each species $s$, calculate the number density $n = \sum_l w_l/\Delta V$, the stress tensor $\Pi_{ij} = \sum_l\gamma_l m_{s,l} w_l v_{l,i} v_{l,j}/\Delta V$, momentum density $U_{p,i} = \sum_l\gamma_l m_{s,l} w_l v_{l,i}/\Delta V$, energy density $U_e = \sum_l\gamma_l w_l m_{s,l} c^2/\Delta V$ and particle flux $F_i = \sum_l w_l v_{l,i}/\Delta V$ at each grid point, where the sum is taken over all (macro)particles in the neighborhood of a cell, $w_l$ is the weight of the macroparticles, $\gamma_l$ is their Lorentz factor, $m_{s,l}$ is the particle mass, $\Delta V$ is the volume element, and $v_{l,i}$ is the $i$-component of the particle's velocity. These lab-frame quantities are outputted automatically in \textsc{Zeltron}.
    \item[2.] Rotate the lab-frame vector and tensor quantities $\vb{U}_p$, $\vb{F}$, $\vb{\Pi}$ to the frame $x',y',z'$ such that the particle flux $\vb{F}'$ points in the $x'$-direction. Note that scalar quantities such as $U_e, n$ are rotationally invariant.
    \item[3.] The components of the comoving thermal pressure tensor in the rotated frame can be obtained by the following equations:
    \begin{align}
        P'_{x'x'} &= \Gamma_b^2\qty(\Pi'_{x'x'} - 2v_b U'_{p,x'} + \qty(v_b/c)^2 U_e), \nonumber\\
        P'_{x'y'} &= \Gamma_b\qty(\Pi'_{x'y'} - v_b U'_{p,y'}), \nonumber\\
        P'_{x'z'} &= \Gamma_b\qty(\Pi'_{x'z'} - v_b U'_{p,z'}), \nonumber\\
        P'_{y'y'} &= \Pi'_{y'y'}, \nonumber\\
        P'_{y'z'} &= \Pi'_{y'z'}, \nonumber\\
        P'_{z'z'} &= \Pi'_{z'z'}, \nonumber
    \end{align}
    where $v_b = F'_{x'}/n$ is the bulk speed of the Eckart (the frame for which particle flux is zero) and $\Gamma_b = (1 - v_b^2/c^2)^{-1/2}$.
    \item[4.] Inverse-rotate $P'_{i'j'}$ to obtain the thermal pressure tensor $P_{ij}$ in the lab frame. The bulk-flow dynamic pressure can be obtained by subtracting $P_{ij}$ from $\Pi_{ij}$ (i.e. $P_{b,ij} = \Pi_{ij} - P_{ij}$). These are then the thermal and dynamic pressure tensors for species~$s$.
    \item[5.] The total bulk-flow kinetic and thermal energies can then be obtained by taking the trace of the respective pressure tensors over all volume $\sum_V\sum_i P_{b,ii}\Delta V$ and $\sum_V\sum_i P_{ii}\Delta V$, for all species present.
\end{enumerate}

The justification for the procedure is as follows. First, note that thermal pressure is defined in the bulk frame of the plasma:
\begin{equation}
    P_{ij} = \int\dd^3\vb{\bar{p}} \bar{f} \bar{\gamma} m_s\bar{v}_i\bar{v}_j, \label{eqn:therm_press}
\end{equation}
with barred quantities denoting comoving-frame quantities and $\bar{f}(\vb{\bar{x}},\vb{\bar{p}},\bar{t})$ is the distribution function satisfying $\bar{n} = \int\dd^3\vb{\bar{p}}\bar{f}$. The comoving frame is connected to the lab (or simulation) frame (denoted by unbarred quantities) by the Lorentz transformation corresponding to the bulk velocity $\vb{v}_b$ and the associated Lorentz factor $\Gamma_b = (1 - v_b^2/c^2)^{-1/2}$. We do not specify $\vb{v}_b$ for now to keep this derivation general. Using the Jacobian of the Lorentz transformation matrix and the fact that particle number is invariant with respect to frame transformation, we have $\dd^3\vb{p}/\gamma=\dd^3\vb{\bar{p}}/\bar{\gamma}$ and $f = \bar{f}$. This gives
\begin{equation}
    \int\dd^3\vb{\bar{p}} \bar{f} \bar{\gamma} m_s\bar{v}_i\bar{v}_j = \int\dd^3\vb{p}\frac{\bar{\gamma}}{\gamma} f \bar{\gamma} m_s \bar{v}_i\bar{v}_j = \int\dd^3\vb{p}\frac{f}{\gamma m_s}\bar{p}_i\bar{p}_j. \label{eqn:therm_pres_transform}
\end{equation}
If we assume the bulk motion is entirely in the $x$-direction, $\vb{v}_b=v_b\vu{x}$, (if it were not, one could simply rotate into a frame in which it is), then Lorentz transformation of the 4-momentum gives $\bar{p}_x=\Gamma_b(p_x - v_bE/c^2)$, $\bar{p}_y=p_y,\bar{p}_z=p_z$. Substituting into eqn.\ref{eqn:therm_pres_transform} and noting that $\Pi_{ij} = \int\dd^3\vb{p} f \gamma m_s v_i v_j$, $U_{p,i} = \int\dd^3\vb{p} f \gamma m_s v_i$, $U_e = \int\dd^3\vb{p} f \gamma m_s c^2$ in the continuous limit gives the equations listed in point 3 of the procedure.
Unlike the prescription described in \citet{Zhdankin-2021}, we have selected the bulk frame to be the zero-particle-flux frame (Eckart's frame), $v_b = F_x/n$.

\bibliography{main}

\end{document}